\documentclass[conference]{IEEEtran}
\IEEEoverridecommandlockouts

\newif\ifpublic
\publicfalse
\publictrue

\usepackage{amsmath,amssymb,amsfonts}
\usepackage{algorithmic}
\usepackage{graphicx}
\usepackage{textcomp}
\usepackage{xcolor}
\usepackage{tabularx}
\usepackage{booktabs}
\usepackage[hyphens]{url}
\usepackage{hyperref}
\usepackage{multirow}
\def\BibTeX{{\rm B\kern-.05em{\sc i\kern-.025em b}\kern-.08em
    T\kern-.1667em\lower.7ex\hbox{E}\kern-.125emX}}

\usepackage[numbers]{natbib}

\input{ryan-meta}
\newcommand{\nAcademic}{341} 
\newcommand{\nMissingAbstract}{4}
\newcommand{\nAbstracts}{337}
\newcommand{\nMissingKeywords}{78}
\newcommand{\nKeywords}{263}

\newcommand{\nAcademicManualOther}{137} 
\newcommand{\nAcademicFAccT}{55}
\newcommand{\nAcademicAIES}{40}

\newcommand{\nAcademicACL}{6} 

\newcommand{\nAcademicOtherACM}{103} 
\newtheorem{definition}{Definition}

\begin{document}

\title{AI auditing: The Broken Bus on the Road to AI Accountability}

\author{\IEEEauthorblockN{Abeba Birhane}
\IEEEauthorblockA{\textit{Mozilla Foundation} and \\
\textit{Sch. of Computer Science and Statistics} \\
\textit{Trinity College Dublin}\\
Dublin, Ireland
}
\and
\IEEEauthorblockN{Ryan Steed}
\IEEEauthorblockA{\textit{Heinz College \& Machine Learning Dept.} \\
\textit{Carnegie Mellon University}\\
Pittsburgh, USA
}
\and
\IEEEauthorblockN{Victor Ojewale}
\IEEEauthorblockA{\textit{Ctr. for Technological Responsibility,}\\ \textit{Reimagination, \& Redesign} \\
\textit{Brown University}\\
Providence, USA
}
\linebreakand
\IEEEauthorblockN{Briana Vecchione}
\IEEEauthorblockA{\textit{Data \& Society} \\
New York City, USA
}
\and
\IEEEauthorblockN{Inioluwa Deborah Raji}
\IEEEauthorblockA{\textit{Mozilla Foundation} and \\
\textit{University of California, Berkeley}\\
Berkeley, USA
}
}

\maketitle

\begin{abstract}
One of the most concrete measures to take towards meaningful AI accountability is to consequentially assess and report the systems’ performance and impact. However, the practical nature of the ``AI audit'' ecosystem is muddled and imprecise, making it difficult to work through various concepts and map out the stakeholders involved in the practice. First, we taxonomize current AI audit practices as completed by regulators, law firms, civil society, journalism, academia, consulting agencies. 
Next, we assess the impact of audits done by stakeholders within each domain. We find that only a subset of AI audit studies translate to 
desired accountability outcomes. We thus assess and isolate practices necessary for effective AI audit results, articulating the observed connections between AI audit design, methodology and institutional context on its effectiveness as a meaningful mechanism for accountability. 
\end{abstract}

\begin{IEEEkeywords}
Evaluation, auditing, accountability, transparency, artificial intelligence, society, law, machine learning, data science
\end{IEEEkeywords}

\section{Introduction}
\label{sec:intro}

The widespread use of artificial intelligence (AI) systems is heavily weighed down by its multitude of related risks. Functional failures~\cite{raji2022fallacy}, disparate performance~\cite{obermeyer2019dissecting,chu2022digital,imana_auditing_2021}, embedded stereotypes~\cite{luccioni_stable_2023,barlas2021see,scheuerman2019computers}, legal incompatibility~\cite{kilovaty2019legally}, privacy violations~\cite{cadwalladr2018revealed}, model inscrutability ~\cite{pasquale2015black} and many more issues plague almost every use case.

Audits are a routine practice in various sectors including finance, public management, healthcare, food safety, and commercial retail~\cite{power_audit_1999, raji_outsider_2022}. The adoption of audits within the AI space, however, is relatively new. 
Despite its imprecision, we use the term ``artificial intelligence'' (AI) to encompass a wide range of deployed products with a significant algorithmic or data-defined component, including but not limited to risk assessments, large base models for computer vision and language, classification models, ``generative'' models, and recommendation systems. 
The inspiration for audit practice in the field of data science, machine learning (ML), and AI 
derives from a variety of related disciplines. Online platform audits, for example, cite social science and critical race  
studies as inspiration~\cite{sandvig_auditing_2014, cherry2018making}, as do several audits of automated decision systems (ADS) and risk assessments~\cite{vecchione_algorithmic_2021}. Some audits of large language models take after security audits~\cite{carlini2023extracting}. Some risk assessment evaluations follow models from traditional experimental design~\cite{lai_towards_2023}, including clinical trials~\cite{rivera2020guidelines}. Meanwhile, internal auditors derive their practice from regulated industries such as finance, aerospace and medical devices~\cite{raji_closing_2020}.

\begin{figure}[!t]
    \centerline{\includegraphics[width=\linewidth]{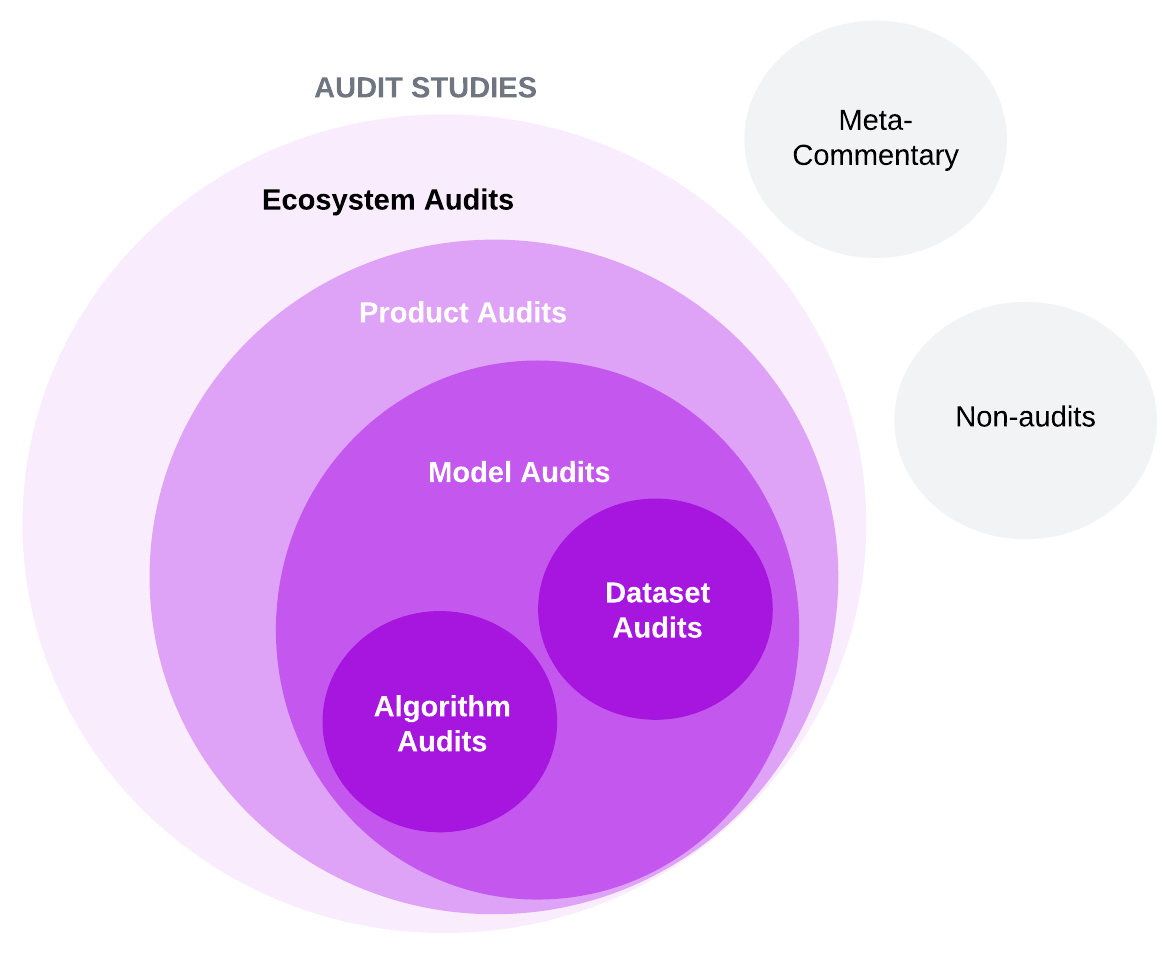}}
    \caption{Audit studies considered in our survey, classified by their scope.}
    \label{fig:taxonomy}
\end{figure}

Across disciplines and contexts, one of the main motivations for conducting audits of AI systems is establishing informed and consequential judgements of the deployed AI systems -- that is algorithmic accountability~\cite{wieringa2020account}.
We thus define an AI audit to be any evaluation of these AI systems, independent of the AI development process, executed for the purpose of accountability~\cite{raji2023change}. 

However, 
unlike other, 
more mature audit industries, AI audit studies do not consistently translate into more concrete objectives to regulate system outcomes, such as influencing voluntary corporate action or internal corporate policies, inciting product recalls, informing product re-designs, as well as shaping broader 
governmental policy and regulation in the form of standards, bans, restrictions or moratoriums of use. In the AI audit context, such impact is fairly uncommon. In this paper, with the aim of enabling auditors and policymakers to achieve accountability outcomes
, we taxonomize audit practices and identify the characteristics of audits that most directly and effectively contribute to audit objectives. 

Some initial work has been done to taxonomize various aspects of AI auditing. Many of these efforts have focused on categorizing \emph{methods}~\cite{sandvig_auditing_2014, vecchione_algorithmic_2021,brown2021algorithm,ada_lovelace_institute_examining_2020,ada_lovelace_institute_technical_2021}, or \emph{types of audit organizations}~\cite{costanza-chock_who_2022, raji_outsider_2022}. Some efforts have also looked at an AI audit's broader institutional~\cite{selbst2021institutional, raji_closing_2020}, policy~\cite{raji_outsider_2022}, and societal context~\cite{vecchione_algorithmic_2021}, through these analyses are often limited to one or at most a handful of case studies. ~\citet{bandy_problematic_2021} is one of few efforts that
systematically classifed audit studies by a broader set of criteria -- audit method, audit target, audit domain and audit objective. However, the study is 
narrow in scope, consisting of
62 exclusively academic audit studies. 

In this paper, we take a much broader and more comprehensive view than past studies. We review the broad audit landscape consisting of academia and six other domains, taxonomizing the key characteristics of their context, goals and practice (Figure~\ref{fig:taxonomy}). Notably, as a mechanism for achieving accountability, we investigate the consequences of these audit investigations and how well the outcomes from the studies match those stated objectives.

\section{Background}

\subsection{\textbf{What is an AI audit?}}

In the context of this paper, we consider the operationalization of audits as a mechanism for accountability in computing --- notably for AI, machine learning, data science, and related fields. We begin by proving 
definitions. 

\begin{definition}
    An \textit{\textbf{audit}} is defined as any \emph{independent assessment} of \emph{an identified audit target} via \emph{an evaluation of articulated expectations} with \emph{the implicit or explicit objective of accountability}.
\end{definition}

Care must be taken to translate this definition in the AI context. 
By \emph{independent assessment}, we mean any measurement done by an entity operationally distinct from the team that engineered the examined AI system. Even if that team is within the same company and composed of corporate employees, the examination only counts as an audit if the auditors are adequately separate from those that built the system, or the audit process itself is distinct from the engineering process for the AI system~\citep{raji_closing_2020, raji_outsider_2022}. 
By \emph{an identified audit target}, we look to studies and investigations that name a concrete and non-abstract, specific object of examination. Ideally, this target is connected to a real-world AI deployment, though sometimes a widely used open-source algorithm or dataset can operate as a stand-in or proxy. For instance,~\citet{steed2021image} looked at bias in an open-source image generation model, reflecting issues later discovered in similar commercial products~\citep{heikkila_how_2022}; and an audit of Proctorio was conducted on the open source model OpenCV on which the commercial product is built~\citep{feathers_proctorio_2021}. Due to a lack of training data disclosure for most AI products (even supposedly ``open source'' models~\citep{touvron2023llama}), many data audits investigate open source datasets~\citep{birhane2021large, solon2019facial, reisner2023revealed} and attempt to generalize conclusions on broader industry practice. Studies of prototypical algorithms (hypothetical models trained by the authors) without concrete, specific targets were not considered.

To be an \emph{evaluation with the objective of accountability}, the audit must incorporate some implied or explicit objective to have the assessment play some role informing consequential judgements about the technology being examined. In order for these judgements to be more concrete, there needs to be some measurement between the reality of the AI deployment and articulated expectations held about a particular deployment.  
    Many academic papers that we examined, for example, studying fairness, performance or safety in an abstract manner~\cite{dwork2012fairness}, without connecting it to anticipated accountability outcome, even implicitly, were not considered by us to be audits~\cite{raji_actionable_2022}.
    Note that the actual \emph{type} of audit target or \emph{criteria of assessment} are not part of the definition of what constitutes an audit. AI audits can thus encompass a wide range of targets, including automated decision systems (ADS)~\cite{angwin_machine_2016, chouldechova2018case}, recommendation systems underlying online platforms or apps ~\cite{chen2015peeking,chen2016empirical}, large base models in computer vision ~\cite{steed2021image,buolamwini_gender_2018}, speech~\cite{koenecke2020racial}, text-based natural language processing~\cite{carlini2023extracting}, or multimodal models~\citep{mandal2023gender}. At times, the evaluations involve domain-specific considerations in hiring~\cite{wilson2021building}, healthcare~\cite{obermeyer2019dissecting}, criminal justice~\cite{lum2016predict} or social service delivery~\cite{koenecke2023popular}. The expectations articulated for these systems can also vary in concreteness and specificity. For instance, some conduct audits specifically for legal compliance~\cite{sapiezynski_algorithms_2022} while others declare expectations more normatively~\cite{raji_actionable_2022}. Others yet are not explicitly labelled as audit work yet satisfy our audit criteria and result in immense explicit and gradual structural change~\citep{noble_algorithms_2018,eubanks_automating_2018}. As a result, we can observe audits that can evaluate and diagnose for a range of performance, safety concerns, as well as broader societal injustices and are 
    not limited to fairness.

\subsection{\textbf{Who conducts AI audits?}}
\label{subsec:who_conducts}

There is a wide range of possible audit practitioners that participate in the audit process~\cite{costanza-chock_who_2022},  
often described as below. 

\begin{definition}
    An \emph{internal auditor} is an entity executing an audit or investigation with some contractual relationship with the audit target~\cite{raji_outsider_2022}. They typically seek to minimize corporate liability and test for compliance to corporate or industry-wide expectations~\cite{raji_closing_2020}. In policy, internal auditors are typically those designated to carry out mandatory corporate audit requirements (e.g. the ``independent auditors'' in Article 37 of the Digital Services Act).
\end{definition}

\begin{definition}
    An \emph{external auditor} is an entity executing an audit or investigation without any contractual relationship with the audit target~\cite{raji_outsider_2022}. They typically execute audits voluntarily  with a broader mandate of identifying and minimizing the harm impacting their constituents.
\end{definition}

~\emph{Internal audits} 
 require a contractual relationship with the audit target. Internal audits are typically conducted by an organization hired by the audit target voluntarily or to maintain compliance with a required legal audit mandate. The auditor in these contexts are hired to operate in a professional capacity to audit the target. 
This often means they are selected and paid by the audit target, though that is not always necessarily the case (e.g., auditors selected and paid by the government). 
 These are the auditors typically referenced in audit mandates. As the executors of more formal audit requirements, these auditors are ideally certified or otherwise qualified~\citep{raji_outsider_2022}, and subject to some form of external oversight and quality control, including but not limited to auditor conduct and reporting standards. 

~\emph{External auditors} typically conduct audits voluntarily by organizations, typically with a broader mandate of research or advocacy. These auditors are not assigned an audit target and do not execute audits on behalf of the audit target but choose to study systems based on the needs and concerns of their constituents. Without any formal connection to the audit target, these auditors typically struggle to access the information necessary to conduct a thorough investigation. Furthermore, without any form of legal protection, data access regime, or oversight, these 
auditors tend to operate quite independently, taking on whatever methods suit their 
objectives. These auditors can be vulnerable to corporate retaliation and methodological skepticism after audit results are released.

\subsection{\textbf{How are AI audits conducted?}}

Although the specific methods used in the execution of AI audits vary widely, 
some terminology are 
deployed regularly to 
describe how audits are executed. Audits can be conducted \emph{ex ante} (before deployment), \emph{in media res} (during an iterative process of design and restricted re-deployment) or \emph{ex post} (after widespread adoption and use). Furthermore, although the details of the AI audit process varies 
widely, the high level structure of that process follows similar stages, which we describe using the taxonomy from \citep{raji_towards_2024}. Note that not all audits follow all these processes.

\subsubsection{Harms Discovery} This is the stage of discovering what to audit for. 
    It involves identifying the audit target entity, 
    possible targeted populations, and the anticipated form of harm or measurement required for a meaningful audit. This can happen, for example,  via direct reporting from the impacted population~\cite{mcgregor2021preventing}, active investigation from the auditors, or other methods~\cite{fink2018opening}.
\subsubsection{Standards Identification} 
    This stage is about effectively articulating the requirements for an ideal AI audit outcome, by naming the standards the auditors will be holding the target to in the evaluation process. These expectations can be as vague as a set of named AI principles~\cite{jobin2019global} or as precise as a specific threshold of performance (e.g., AI hiring tools' adherence to the 4/5ths rule~\cite{ajunwa2016hiring}).
\subsubsection{Performance Analysis} The core of the audit is the actual evaluation itself. There are a wide diversity of methods available to inform final assessments, ranging from qualitative to quantitative approaches~\cite{bandy2021problematic}. Each method involves different degrees of complexity and challenges tied to data acquisition, model access, and measurement. 
 \subsubsection{Audit Communication and Advocacy} Following the evaluation of the AI system, there is often also some activity to disseminate and translate audit results to some relevant stakeholders. That audience could include the audit target, regulators and the public ~\citep{raji_actionable_2022, krafft2021action}---but could also include internal stakeholders within the organization such as a legal, business, product or engineering team~\citep{raji_closing_2020}.
  
\section{Methods} 
In order to review the AI audit research and practice landscape, 
we conducted a wide-ranging literature review of academic 
work published in a range of venues. We reviewed websites, reports, and relevant documentations from numerous non-academic audit practitioners. 
For academic research, 
we collected a comprehensive sample of $N=\nAcademic$ audit studies published in interdisciplinary computing and related venues from 2018--2022. Figure~\ref{fig:bar-time} shows the number of collected academic studies published per year, grouped by audit label type, while Table~\ref{tab:academic-lit} describes our search method for each source and lists the number of identified pieces of literature collected for each. For audit practices outside academia, we identified major domains that are both relatively established in existing academic literature and currently emerging as key players in the AI audit ecosystem (inspired by the main categories identified by~\citep{costanza-chock_who_2022}): journalism, civil society, government, consulting agencies and corporate audits, 
and legal firms. We collected specific cases 
that serve as concrete examples of each domain. 
These methodological choices not only facilitate coherence with existing scholarship but also enhance the reliability and comparability of our findings within broader academic discourse on the subject.

Our list is not exhaustive of all academic audit literature or audit domains, but it provides the most comprehensive sample (to our knowledge) of the current state of AI audit ecosystem, encompassing from academic research to practice and everything in-between. 
The full list of audit studies we found using these methods is included in the supplementary information (see Appendix~\ref{app:supplementary}).

\subsection{\textbf{Academic literature}}
Audits in the academic context, conducted by authors from universities, non-profits, and tech companies, encompass a wide variety of disciplines, methods, and aims. Academic audits are published in various formats and venues, for example, as books, journals, and conference proceedings. Over the past five years, interdisciplinary computing conferences have become a central place where work around the topics of fairness, accountability, and transparency is discussed and published, especially within 
the ACM conferences, FAccT and AIES, 
two emerging main conferences in the space. In this regard, conference proceedings are not only the most common way of sharing work amongst the AI ethics (broadly defined) community, but such formats of publication also share commonalities such as relatively standardised style and presentation. We have, therefore, selected conference proceedings as a primary focus for our systematic reviews of academic audits. 

Our search is not exhaustive---it did not, for example, include potential audit work that might have been covered in high-impact general audience journals such as Science, Nature, or PNAS. We also did not consider work published in non-interdisciplinary social science venues---in particular, work from economics or sociology. 
Instead, our sample represents the kinds of audit work recently published at interdisciplinary computing and related conferences.

\subsubsection{Searching for audit studies} 
\begin{figure}[!t]
    \centerline{\includegraphics[width=\linewidth]{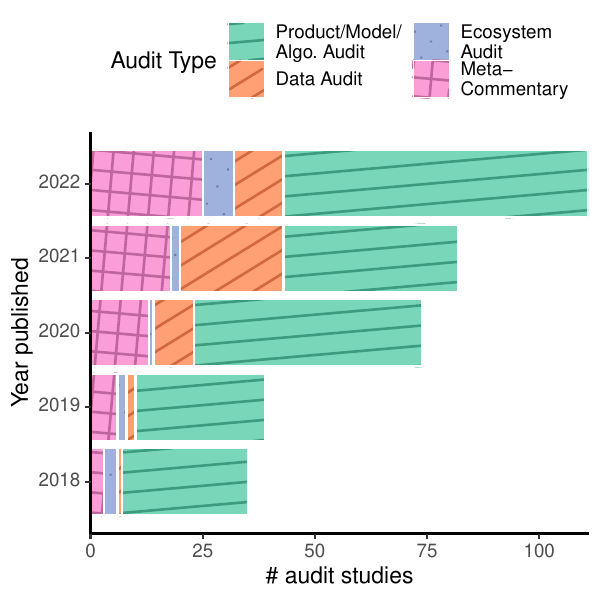}}
    \caption{Number of collected academic audit studies published in each year, grouped by manually labelled audit type. Studies that involve multiple audit types are counted only under the best-fitting category.}
    \label{fig:bar-time}
\end{figure}

We found $N=\nAcademic$ academic papers that met our criteria for an audit study (Figure~\ref{fig:bar-time}, Table~\ref{tab:academic-lit}). We identified academic audit studies with two methods. First, we manually reviewed the conference proceedings from 2018 to 2022 for a selection of smaller interdisciplinary computing conferences: Fairness, Accountability, and Transparency (FAccT), Artificial Intelligence, Ethics, and Society (AIES), Equity and Access in Algorithms, Mechanisms, and Optimization (EAAMO), Association for the Advancement of Artificial Intelligence (AAAI), Computer-Supported Cooperative Work \& Social Computing (CSCW), International Conference on Computational Social Science (IC2S2), and the International World Wide Web Conference (WWW). 
We selected these conferences as venues where audit work is likely to appear, but still, audits are not the primary focus of these conferences. Of the 5 years of proceedings we reviewed, only $N=237$ papers met our criteria for an audit study.

Second, to expand our search to other, larger conferences, we automated the inclusion process: we applied keyword searches to both the ACM Digital Library (excluding FAccT, AIES, and CSCW proceedings, which we reviewed manually) and the ACL Anthology. In these two libraries, we searched for papers published from 2018 to 2022 with one or more of six keywords in the title: ``audit'', ``accountability'', ``case study'', ``bias'', ``fairness'', and ``assurance''. For ACL, where the searches were web-based, we manually reviewed the search results for audit studies online, downloading only the few studies that met our criteria ($N=\nAcademicACL$). For ACM, we downloaded the search results. There were thousands of papers in the queries in total. Three of the keywords yielded relatively few ACM entries: 
``audit'' ($N=39$), ``accountability'' ($N=87$), and ``assurance'' ($N=65$), after removing duplicates. For the rest of the keywords, which returned over 500 results each, we narrowed our manual search by considering only the first 500 ``most relevant'' entries returned by the ACM search engine. After removing duplicates and other incomplete entries, each keyword yielded the following number of pieces of literature: ``fairness'' ($N=404$), ``bias'' ($N=318$), and ``case study'' ($N=472$). We further excluded everything that is not a standard academic paper (such as panels, abstracts, workshops, and tutorials) for consistency. We then manually reviewed each abstract and, as before, excluded those that did not meet our criteria for an audit study, leaving a total of $N=\nAcademicOtherACM$ audit studies from ACM conferences other than FAccT and AIES.

\subsubsection{Classifying audit types} We classified all the audit studies we collected into two main categories: audit studies and meta-commentary (depicted in Figure~\ref{fig:taxonomy}). We further classified audit studies into two kinds: \emph{product/model/algorithm} audit studies
and \emph{data} audit studies. Some studies targeted specific products (e.g., YouTube comment moderation \citep{jiang_bias_2019}); others targeted only the AI models behind the products (e.g., an investigation of tax audit models used by the IRS \citep{black_algorithmic_2022}); still others targeted only the algorithms used to build the model (e.g., pre-trained large language models such as GPT-3 \citep{abid_persistent_2021}). Since the methodologies for these kinds of audits are similar, we classify them together. We distinguish these studies from \emph{data} audit studies---also a kind of model audit---which specifically target training or benchmark datasets (e.g., ImageNet \citep{birhane_large_2021}) and tend to use different methodologies.

During the iterative classification process, an additional category emerged: \emph{ecosystem} audit studies. Ecosystem audit 
go beyond datasets, models, and products and examine communities and sociotechnical environments (digital or physical) impacted by or are critical components to an AI system's operation. \citet{brown_toward_2019}, for example, audited the impact of child welfare service algorithms by conducting interactive workshops with front-line service providers  and families affected 
by such algorithms. \emph{Meta-commentary} studies are those 
that examine auditing as a practice, including 
studies that put 
forward novel mechanisms and processes as a viable \emph{method} for auditing certain systems as well as those proposing to  
improve 
existing audit processes. Meta-commentary also includes \emph{critiques}, work that interrogates  
the effectiveness and merit of auditing as a practice. \todo[examples of each from spreadsheet]

Note that not all papers fit neatly into one of the above categories, and sometimes we find some papers incorporating elements of two---and, rarely, three---categories. For example, methodology papers often also 
use a specific case study to illustrate the method in practice. In our 
analysis (Fig.~\ref{fig:bar-time}), we consider only the classification that best fits the paper.

\begin{table}
\centering
\caption{Audit literature from academic sources}
\label{tab:academic-lit}
\begin{tabularx}{\linewidth}{ p{11em} X c}
\toprule
\textbf{Source} & \textbf{Search method} & \textbf{$N$} \\
\midrule
FAccT & Manual review of proceedings — titles and abstracts, then full paper where classification still unclear & {\nAcademicFAccT} \\

AIES & '' & {\nAcademicAIES} \\
EAAMO, AAAI, IC2S2, and WWW & '' & {\nAcademicManualOther} \\

\addlinespace[0.25em]\hline\addlinespace[0.25em]

Other ACM proceedings & Keyword filters, then manual review & {\nAcademicOtherACM} \\

ACL Anthology & '' & {\nAcademicACL} \\

 & \textbf{Total} & {\nAcademic} \\

\bottomrule
\end{tabularx}
\end{table}

\subsection{\textbf{Non-academic domains}}
For audit work outside the academic domain, we identified the following categories where audits are regularly conducted and becoming an established practice: journalism, civil society, law firms, regulatory audits, and consulting agencies and corporate audits. 
For each category, we gathered sources, often information on auditors websites, audit reports, and other relevant documentation produced by various bodies (whenever they are available), which we reference in our analysis (\S\ref{sec:other-domains}). Based on information from these sources, for each category we identified a number of institutions, organisations, agencies, or audit reports that served as concrete examples of the given category. Appendix~\ref{app:supplementary} lists all the reports, webpages, and other documents we reviewed in each audit category and institution.

\subsection{\textbf{Analysis}}
For both the academic audits and each non-academic audit source, we reviewed published audit studies, documents, and public webpages to summarize information on the following:
\subsubsection{Context} Context includes the specific structural factors shaping the audit (all as stated by the auditor): what motivated the audit (\emph{Motivations}), its 
goals (\emph{Goals}), the main artefact or system under investigation (\emph{Audit Target}), the specific harms or concerns investigated (\emph{Types of Harms}), and ways in which the audit may or may not 
shift power between stakeholders~\citep{kalluri2020don} (\emph{Institutional Context}). 
\subsubsection{Methodology} Methodology includes the main techniques and procedures used to investigate the target artefact---for example, quantitative evaluation (typical in academic audit studies) or forensic analysis or qualitative interviews (found more often in civil society audits).
\subsubsection{Impact} Impact refers to changes that occur to the target artefact, the target audit, or the institutional environment as a direct consequence of the audit. Impacts could include policy developments, alterations to an algorithm, or monetary fines for certain violations.
While impacts are well documented in some domains (e.g., journalism), the impact of an academic audit, for example, is not often 
clear at the time an audit is published.
Detecting, measuring, and quantifying 
impact is challenging. Subsequently, 
we were as 
inclusive as possible of many different kinds of impacts, e.g., inspiring media coverage of an overlooked harm~\cite{metcalf_algorithmic_2021}. We considered notable documented evidence  
in the audit reports themselves or from related news stories, whenever available. 
In our analysis, we make a note wherever documented impacts were unstated and unclear (in academia, for example).

For nearly all the academic studies we found, 
our dataset includes abstracts ($N=\nAbstracts$) and author-selected keywords ($N=\nKeywords$). In order to supplement our qualitative findings 
for academia (\S\ref{sec:academic}) with quantitative statistics, we analysed 
key terms (e.g., ``accountability'') that were used most frequently by authors. Appendix~\ref{app:keyword} describes how we selected key terms related to the criteria above.

\section{Results: Auditing in Academia}
\label{sec:academic}
\todo[Add more citations from spreadsheet?]

\subsection{\textbf{Audit types}}
\label{sec:case-studies}
\subsubsection{Product/model/algorithm audits}
Most audit work fell under the category of product-level case studies. These audit studies target a mix of social media platforms, algorithms for administering public services, large language and vision models, and search engines. 
These case studies 
typically evaluate specific deployed systems.
These case studies mainly diagnose failures, errors, disparities. For example, some variation of ``bias'' is mentioned in 32.5\%, and ``fairness'' 
in 21.2\% of abstracts (see Table~\ref{tab:kwds-motivation-pivot}). Less commonly, these studies call for model builders and practitioners to make amendments accordingly. For example, the term  
``accountability'' appeared 
in only 
14\% of abstracts, less often than in ecosystem audits (33.3\%) or meta-commentary (28.1\%).

\subsubsection{Data audits}
Data audits typically focus on evaluating specific datasets, often 
targeting datasets used to train large models \citep{birhane_multimodal_2021} and 
sometimes interrogate the benchmark datasets used for model evaluation, such as COMPAS \citep{bao_its_2021}. 
Oftentimes, these studies 
emphasise (both implied and explicit) shifting  
norms around data use and benchmarking practices, while 
fewer explicitly emphasize  
holding dataset creators accountable. For example, ``accountability'' is mentioned in 
9\% of data audit abstracts, 
while ``bias'' is mentioned in 39.1\% of abstracts (more in this category than any other). These studies typically focus on surfacing harms ranging from representation and stereotyping to privacy and, more recently, copyright protection~\cite{samuelson2023generative}. The type of methods used include quantitative measurement (such as 
the incidence of NSFW images), simulation, ablation (removing or changing certain aspects of the dataset and measuring the result), and critical assessment. Like case studies, data audits tend to be conducted by a range of academic, non-profit, and corporate authors examining open-sourced and academic datasets.

\subsubsection{Ecosystem audits} These studies target public services ---predictive risk models used by child welfare agencies, for example ~\citep{stapleton2022imagining, brown_toward_2019}---more often than any other kinds of audits. Abstracts of these studies also mention specific domains such as hiring and education more often than any other audit type (Table~\ref{tab:kwds-targets-pivot}).
Harms and concerns are often concretely defined 
and audit are typically carried out with the expectation of 
subsequent change both to specific stakeholders and sometimes, 
society at large. 
The term ``accountability'', for example, is mentioned in over 33\% of ecosystem audit abstracts, the highest of any category. is goal, 
These studies often utilise a wider range of methods, including qualitative interviews, surveys, workshops, and literature reviews \citep{radiya-dixit_sociotechnical_2023, stapleton_imagining_2022, woodruff_qualitative_2018}. Given the scope of investigation, methods are often all-encompassing and diffuse, less precise than more scoped audit investigations~\cite{solaiman_evaluating_2023}, where they 
are much more likely to employ qualitative and participatory methods than other audit studies. 6 out of 15 ecosystem audit abstracts, for example, mention ethnography, interviews, workshops, or other qualitative methods, while  
two explicitly 
use the term ``participatory'' (see Table~\ref{tab:kwds-methods-pivot}).

\subsubsection{Meta-commentary \& critique}
We found that 28.1\% of meta-commentaries mention ``accountability'' in the abstract. 
The type of methods they used 
include surveys, interviews, and literature reviews. These kinds of studies were nearly exclusively conducted by academic, non-profit, or government authors---for example, 
guidance from government agencies such as NIST's AI Risk Management Framework \citep{tabassi_ai_2023}.

Many meta-commentary papers 
aimed  
to develop 
a methodology and many  
considered the development of rigorous audit methods 
as a vital contribution. Audit methods these studies put forward include both  
qualitative and quantitative approaches.  
These studies also examine audit practitioners themselves, interrogating norms around algorithm evaluation and audit practice. The harms at issue tend to be less well-specified; most mention some form of independence, fairness, privacy, or recourse. These studies 
rarely mention 
affected stakeholders in method design. Similarly, a couple of tool development papers developed standardised tools, usually quantitative, for auditing or for algorithmic recourse. 

\subsection{\textbf{Impact by audit type}} 

\subsubsection{Product/model/algorithm audits}
We found 
few audit studies that acknowledge and address power asymmetries. 
Many prominent papers we surveyed are collaborations between authors from academic institutions and 
large tech corporations, usually examining their own systems and datasets, and rarely explicitly calling for systemic change, auditing 
systems, 
often without involvement from affected stakeholders. 
Unlike audit work by journalists and 
civil society, 
academic audit work tends to follow academic norms where the objective is academic publication. Subsequently, more often than not, audits are seen as an academic, intellectual exercise than practices directly linked to real world consequences. 
Audit findings are rarely presented with 
demands for concrete systemic change. This does not, however, mean that academic audits are not impactful. To the contrary, seminal audit case studies such as Gender Shades~\citep{buolamwini_gender_2018} have not only resulted in meaningful improvements to deployed systems~\citep{raji_actionable_2022} but also have come to establish algorithmic audits as a field of enquiry. The impact of such work is, however, rarely obvious at the time of publication but becomes apparent gradually over time.  
For industry collaborations, the audits may result in organisational reform. For example, case studies co-authored by authors with big tech affiliations such as~\citep{ramesh_how_2022} may have contributed to tightening Google Play Store app data access restrictions~\citep{singh_google_2023}. Generally, 
academic audit results that publicly call out audit targets~\citep{raji_actionable_2022}, that tend to be picked up by news outlets such as MIT Tech Review, activists, and regulators tend to bring about the most observable 
changes. 

\subsubsection{Data audits}
The impact of data audits are also often unclear, though prominent data audits sometimes resulted in changes to benchmarks — an audit of 80 Million Tiny Images, for example, resulted in the withdrawal~\citep{torralba_80_2020} of the dataset~\citep{birhane_large_2021} 
and a Financial Times investigation of Microsoft’s dataset of ``celebrity'' faces even resulted in its discontinuation \citep{murgia_whos_2019, murgia_microsoft_2019}.

\subsubsection{Ecosystem audits}
Ecosystem audits often reveal 
comprehensive institutional or legislative policy demands in advocacy. This includes,  connecting performance and privacy concerns involved in facial recognition use by 
law enforcement~\cite{radiya-dixit_sociotechnical_2023}, breaking down sources of bias throughout the model development cycle~\cite{suresh_framework_2021}, systematic environmental costs~\cite{schwartz_green_2020, rakova2023algorithms} or the labor issues across the entire supply chain of AI development~\cite{crawford2021atlas}. 

\subsubsection{Meta-commentary \& Critique}
Although the immediate impact is directly quantifiable, 
meta-commentaries provide important mechanisms that allow audit studies and practices 
to zoom out, self-reflect, and evaluate the overall picture and direction, all of which is a crucial element for grounding audits in concrete foundations and optimal accountability. 

More specifically, a certain type of meta-commentary that seems to be of particular importance are critique studies. This category of work engages in  
reflexive commentary critiques of auditing as a practice. A few pieces of critical work interrogate the effectiveness, shortcomings, and limitations of audits. Such work highlights structural issues such as historical power asymmetries  
that might be reinforced by common academic auditing practices, or how audits might serve as a smoke-screen for corporate responsibility, such as, audit washing~\citep{gansky_counterfacctual_2022,goodman_ai_2022} or how audits might lead to simplistic technological solutionism \citep{sloane_silicon_2022}. They draw on qualitative interviews, literature reviews, and statistical analysis. Like meta-commentaries, the impacts of these works are not easily measured, though many of these studies are highly cited.

\section{Results: Auditing Outside Academia}
\label{sec:other-domains}

In this section, we review a sample of audits and audit practices from outside academia --- law firms, consulting agencies and corporate audits, journalism, civil society, government, and civil society  --- 
examining how factors such as institutional context affects the practice of auditing and accountability.  
These auditors operate in a variety of domains and deploy various methods throughout the audit design, development, and execution process. For each domain, we pay particular attention to the details of the audit \emph{context} and \emph{methodology}, then connect this to the observed \emph{impact} derived from  
audits for that particular domain. Our analysis criteria and results are summarised in Table~\ref{tab:summary} (The full list of documents we analysed is found in Appendix~\ref{app:supplementary}). 

\begin{table*}
\caption{Summary of analysis across AI auditing domains.}
\label{tab:summary}
\begin{tabularx}{\linewidth}{ p{5em} p{8em} X p{10em} p{10em} X X }
\toprule
\multirow{2}{*}{\textbf{Domain}} & \multicolumn{4}{c}{\textbf{Context}} & \multirow{2}{*}{\textbf{Methods}} & \multirow{2}{*}{\textbf{Impacts}} \\
 & \emph{Motivation} & \emph{Target} & \emph{Types of Harms} & \emph{Inst. Context} &  & \\ 
\midrule
\textsc{Academia} & Accountability, make policy, establish standards & ADS, 
online platforms, training datasets, audit practitioners & Functionality failures, 
disparate impact, 
privacy harms, stereotyping, 
manipulation, ethics washing, unfairness & 
Academic, non-profit, \& corp. authors studying systems built by same, occasionally involving stakeholders & interview, survey, simulation, quant \& qual eval, theory,  lit review, 
  framework devt, ablation & Often unclear; media coverage; occasionally, reforms to target systems \\ 
\addlinespace[0.25em]\hline\hline\addlinespace[0.25em]
\textsc{Civil Society} & Accountability, advocacy, make policy & ADS, 
online platforms, surveillance tech, biometric data use & Fair use harms, privacy harms, censorship, disparate impact, HR violations, deceptive claims, hate speech & Non-profit, independents usually advocating on behalf of harmed stakeholders &  Activism, interviews, visualisation/media, forensics, 
public records requests & Civil suits won, media coverage, moratoriums, abolition \\ 
\addlinespace[0.25em]\hline\addlinespace[0.25em]
\textsc{Journalism} & Accountability & Online platforms, large tech companies, ADS (public services) & disparate impact, fraud, non-compliance, privacy harms, cultural genocide & Targets mostly tech firms \& gov’t agencies on behalf of
/in consultation with 
public; funded by foundt'ns/private donors & Investigative reporting, quant. eval., interviews, data donation, scraping, docs review & Regulatory action, civil suits, reforms to target systems, inspiration for research \\ 
\addlinespace[0.25em]\hline\addlinespace[0.25em]
\textsc{Government} & Enforce regulations, establish standards & Data sharing and management; surveillance tech, ADS & Non-compliance, privacy harms, racial disparities, physical safety, functionality & Targets companies and gov’t agencies; may or may not have enforcement authority & Interviews, testing, docs review, quant. eval. & Monetary penalties, standard-setting \\ 
\addlinespace[0.25em]\hline\addlinespace[0.25em]
\textsc{Corporate Audit Reports} & Identify ethical issues & Internal APIs, platforms & Human rights “impacts”, racial injustice, voter suppression, hate speech & ``Voluntary'' reforms, often at request of \& in consultation with external stakeholders & Stakeholder/expert consultation, interviews & Reforms to practices, commitments to future changes \\ 
\addlinespace[0.25em]\hline\addlinespace[0.25em]
\textsc{Law Firms} & Compliance, impact assessment (BNH.AI, AWO); accountability (Foxglove) & Public services, online platforms, gig platforms, other private companies & Non-compliance, injustice/disparate impact, privacy harms, safety, digital manipulation & 
For-profits and public services (BNH.AI, AWO); consulting for harmed  stakeholders (Foxglove) & Testimony, docs review, policy development, legal research & Reform, revision, abolition, 
forced disclosure of secret contracts,  
 (Foxglove) \\ 
\addlinespace[0.25em]\hline\addlinespace[0.25em]
\textsc{Consulting Agencies} & Compliance, impact assessment, mitigate financial/PR risk & ADS, 
online platforms, gig work platforms, surveillance tech & Noncompliance, disparate impact, non-functionality, privacy harms, security, risk & Consulting for for-profit audit targets; may consult stakeholders (Babl AI) & Assessment frameworks, docs review, data governance planning 
& Set precedent for policy (ORCAA); otherwise not clear or unstated \\
\bottomrule
\multicolumn{7}{p{0.98\hsize}}{\footnotesize ADS: automated decision systems.}
\end{tabularx}
\end{table*}

\subsection{\textbf{Law Firms}}
There is a large industry devoted to data governance legal services, but firms offering legal services for AI auditing are less common.
Three such merging boutique law firms  
are: Luminos.Law~\citep{luminoslaw_public_nodate} (formerly called BHN.AI), Foxglove~\citep{foxglove_who_nodate}, and AWO~\citep{awo_awo_2023}. These firms represent three different institutional arrangements in audit-related legal services.

\paragraph{Context} 

Law firms operate as both \emph{internal} and \emph{external} auditors. They 
can be hired by the audit target to conduct an internal assessment as part of a regulatory requirement or in legal defense. Law firms also work with  
representatives of impacted populations (often pro bono) to externally investigate an audit target to collect material evidence of perceived or reported harms. Both scenarios are typically prompted by individual or collective complaints post-deployment.

The organisational structure, business model, and subsequent selection of clientele determines the general objectives for these three organisations. Foxglove---a non-profit organisation advocating for 
the least empowered---aims to hold AI operators to account, aspiring to ``stand up to tech giants and governments'' and ``make tech fair for everyone'' \citep{foxglove_who_nodate}. Accordingly, the main audit target for Foxglove includes data practices of social media platforms, governmental institutions, as well as the working  conditions of content moderators, warehouse workers, and gig workers.

Luminos.Law and AWO---both for-profit firms---offer compliance and public policy services. They carry out audit work at the behest of their clients, with the goal of ensuring compliance with standards, data rights, data protection due diligence, and regulatory obligations. Luminos.Law's client testimonials feature Fortune 100 \& 500 tech firms \citep{luminoslaw_clients_nodate}; AWO's ``commercial practice is balanced with giving those less-resourced a voice'', according to the firm, and their client testimonials feature many non-profit and university clients \citep{awo_services_nodate}. The type of harm and concerns that underlie these firms' services vary accordingly. For Luminos.Law, these are assessment and assurance around legal compliance, legal defence and liability \citep{luminoslaw_our_nodate}. Similarly, AWO focuses on auditing for privacy and data rights, safety, surveillance violations, digital manipulation and exploitation \citep{awo_services_nodate, awo_blog_nodate}. The types of harm and concerns that drive audit practice for Foxglove, on the other hand, include violations of justice, disparate performance \citep{dark_home_2020, dark_we_2020}, and specific breaches of data governance law \citep{foxglove_areas_nodate, hegarty_government_2021}.

We see a stark difference in power asymmetries amongst these three organisations. As for profit forms, both Luminos.Law and AWO's objectives, missions, and practices are shaped by their business models that prioritise the needs of their clients, which are often wealthy and powerful corporations. Foxglove's work 
directly or indirectly pushes to shift power from the most to the least powerful. Subsequently, Foxglove frequently works with groups such as content moderators, warehouse workers and gig-workers, groups that are often underpaid, over-exploited, and disfranchised.

\paragraph{Methodology}
Foxglove's cases often involve specific campaigns, petitions, and/or case studies. Foxglove uses methods such as legal compliance analysis, interviewing, and anecdotal evidence, or leans on previous research to diagnose and highlight harms, concerns, and 
to pursue legal challenges. 
Both Luminos.Law and AWO work within the needs of their clients, which include big tech companies and start-ups. Within such a context, the main audit target for Luminos.Law are models and data, while audit target include social media platforms \citep{tau_banning_2023} and hiring algorithms \citep{lynch_new_2023}, based on media coverage of the firm \citep{luminoslaw_public_nodate}.
AWO similarly carries out documentation analysis, legal compliance analysis, and policy development using previous research as a basis \citep{awo_blog_nodate}.

\paragraph{Impact} As firms that provide audits as a service to clients, some details of their practices are inaccessible---particularly for Luminos.Law---because much of their work is explicitly marketed as ``privileged and confidential'' \citep{luminoslaw_clients_nodate}. Information on the impacts of their commercial work is therefore difficult to identify.   
AWO's blog, for example, mostly features work done in collaboration with civil society organizations like the Ada Lovelace Institute \citep{awo_awo_2023-1}. Luminos's public portfolio includes some standards guidance, a bias calculator for New York's new audit law \citep{luminoslaw_microwave_nodate, cumbo_local_2021}, and an audit of the open-source large language model RoBERTa \citep{brennen_ai_2022, calix_saisiyat_2022}, but the impacts of these audits on practice are not evident.

Foxglove’s work, on the other hand, often results 
in 
significant impactful changes including the reversal of 
Ofqual’s A level grading algorithm~\cite{lee_student-level_2020}, halting the use of visa-streaming algorithm that was deployed by the UK Home Office~\citep{the_joint_council_for_the_welfare_of_immigrants_we_2020}, and forcing disclosure of a secrete contracts between corporations and the UK government; for example, the ``NHS Covid-19 data deals''~\citep{fitzgerald_under_2020} and exposing contracts between Palantir and the UK government. 

\subsection{\textbf{Consulting Agencies \& Corporate Audits}}
Other organizations offer audit services beyond legal and policy advice. A crop of recent startups, such as, 
Arthur AI and Fiddler AI, offer model monitoring services with fairness and privacy components. Others offer consulting specifically, auditing as a service, including boutiques such as, Parity Consulting \citep{parity_consulting_parity_nodate} as well as large consulting firms such as Deloitte \citep{cassidy_auditors_2022}, McKinsey \citep{buehler_identifying_2021}, and Accenture \citep{accenture_ai_2023}. 

Because the product-specific work stemming from an internal audit team is rarely published, corporations  
occasionally publish a report in conjunction with an external consulting agency. For example, Business for Social Responsibility (BSR) has conducted audits on behalf of Google for its celebrity facial recognition system~\cite{business_for_social_responsibility_google_2019}, and  Facebook regarding its human rights impact in Myanmar~\cite{warofka2018independent}. At times these consulting agencies  
are also part of a regulatory process.  
For example, the management consulting firm called Guidehouse Inc. played the role of an ``independent third-party reviewer'' in the Department of Justice (in the US) settlement with Facebook regarding bias in its advertising~\cite{office_of_public_affairs_justice_2022}. 

Here, we examine three consulting agencies that have been involved with prominent audit case studies: ORCAA~\citep{orcaa_orcaa_2023}, Eticas~\citep{eticas_algorithmic_2022}, and BABL AI~\citep{babl_ai_about_nodate}. 

\paragraph{Context} All three agencies provide \emph{internal} audits as a service with the main objective of algorithmic accountability, and 
engage primarily in case-by-case consulting with private and public sector clients such as HireVue, Airbnb, Proctorio \citep{babl_ai_about_nodate}, the states of Illinois and Colorado \citep{orcaa_orcaa_2023}, the Allegheny County Health Department \citep{eticas_algorithmic_2022}, the cities of Barcelona \citep{eticas_algorithmic_2022} and Amsterdam \citep{orcaa_orcaa_2023}, and the University of Iowa \citep{babl_ai_about_nodate}.  Within the bounds of client-agency agreement, the main audit targets for these agencies include social media platforms (such as TikTok and YouTube), ADS, FRT, ride hailing apps (such as Uber, Cabify, Bolt), predictive scoring systems, hiring algorithms, and healthcare algorithms~\citep{orcaa_orcaa_2023,eticas_algorithmic_2022,babl_ai_services_nodate}. Typically, the audits are ex post, though some can also be completed pre-deployment (e.g. BSR celebrity facial recognition audit was conducted before model release).  
ORCAA and Eticas also undertake some internal audit methodology development work.

ORCAA conducts audits in order to assess regulatory compliance, performance testing, as well as to measure and mitigate disparate performance \citep{orcaa_orcaa_2023}. BABL AI conducts audits for bias, risk, and impact assessment \citep{babl_ai_services_nodate}. Like Luminos, ORCAA and BABL AI, both offer services to help companies comply legal requirements, for example, in the US, the New York Local Law 144 \citep{cumbo_local_2021}, which requires annual audits of AI hiring tools. Eticas similarly conducts audits to 
ensure security and data protection and to assess issues such as fairness, bias, and model accuracy \citep{eticas_algorithmic_2022}.
The type of concerns and harms these consulting agencies mention in public materials vary. ORCAA, for example, mainly focuses on measuring bias---such as discrimination, race, and gender---and assessing for regulatory compliance \citep{orcaa_orcaa_2023}. Eticas mainly focuses on assessing algorithmic systems, for example, for functionality and detecting unfair practices towards protected groups \citep{eticas_algorithmic_2022}. Similarly, the main types of harm BABL AI focuses on include bias, fairness, transparency as well as assessment for standards and privacy compliance \citep{babl_ai_about_nodate}.

It is difficult to asses how these three consulting agencies might shift power with clarity. 
However, they target discrimination, fairness, and gender and ethnicity disparities and aspire to lofty goals---BABL AI, for example, seeks to ``prioritize human flourishing'' \citep{babl_ai_about_nodate}. But, like the legal consulting firms, these audits are subordinate to client needs, 
and those clients include large corporations, startups, and government bodies. Ultimately, these audits primarily serve those entities---for example, by protecting clients from concerns such as organisational and reputational crisis
---and help clients build trust with stakeholders~\citep{eticas_algorithmic_2022,babl_ai_about_nodate,orcaa_description_2020}. 

\paragraph{Methodology} We have sparse information on audit methodologies used within these agencies. From the information we can gather on their web-pages, these agencies sometimes develop 
internal audit tools. ORCAA, for example uses ``Ethical Matrix Framework'' \citep{orcaa_orcaa_2023}, while BABL AI has developed a set of criteria that can be used to conduct bias testing in their ``process audit'' \citep{babl_ai_services_nodate}. Other methods include reviewing documents, interacting with stakeholders, and ethics due diligence vetting, 
the details of which 
are \textit{not} provided. 

\paragraph{Impact} The impact of these types of audits, more particularly impact as a direct consequence of these three agencies, is not clear, particularly, with Eticas and BABL AI. ORCAA’s work has seen some impact (albeit as an indirect influence) on policy on the US White House AI Ethics Blueprint~\citep{samuel_theres_2022}. 
ORCAA also conducted an audit of HireVue's early career and campus hire assessment tools \citep{orcaa_description_2020, engler_independent_2021}. HireVue subsequently declared its intention to make changes to its practices (but only following a formal complaint from the Electronic Privacy Information Center to the U.S. FTC 
\citep{the_electronic_privacy_information_center_epic_complaint_2019}), including halting the use of facial analysis in its tools, but received criticism for misrepresenting ORCAA's analysis and not taking bolder steps \citep{engler_independent_2021}. %
HireVue also commissioned the audit and defined both the scope of evaluation and the extent of its impact---a problem of independence raised frequently in prior work \citep{raji_outsider_2022, engler_independent_2021, goodman_ai_2022, costanza-chock_who_2022}.

\subsection{\textbf{Journalism}}
Journalists have a long history of investigating and reporting on AI systems~\cite{diakopoulos2015algorithmic}. 
Investigative journalists at the Wall Street Journal, the New York Times, the MIT Technology Review, and many other outlets have unearthed concrete examples of systematic AI harms \citep{hao_cleaning_2023, hill_wrongfully_2020, hao_south_2022, hao_how_2022, heikkila_viral_2022}.

We examined work from two of the most prominent outlets that have conducted extensive AI audits: ProPublica and The Markup. ProPublica~\citep{propublica_investigative_2023} carried out 
a foundational investigation into criminal risk assessment in 2016, that has set precedence for the field of AI auditing 
\citep{angwin_machine_2016}. The Markup~\citep{the_markup_about_nodate} is a relatively newer outlet focused on data-driven investigations. 

\paragraph{Context} Audits from both The Markup and ProPublica tend to be~\emph{external} 
focusing on specific types of targets. These include, AI systems that are commonly used by large corporations or social media platforms for advertising, hiring, and ranking; government 
ADS, or public programs with digital sites; as well as content moderation systems and related labor conditions. Both organisations  carry out audits with the stated objective of accountability. The Markup's slogan reads, ``Big Tech is watching you. We're watching big tech'' \citep{the_markup_about_nodate}. The type of harms and concerns both organisations investigate, surface, diagnose, and evaluate include disparities in performance (for instance, along the dimensions of gender, race, ethnicity, age), injustice, discrimination, privacy violations, fraud, and legal compliance breaches. As harm discovery happens through publicly submitted journalistic ``tips'', the audits are typically ex-post.  

\paragraph{Methodology} Both Propublica and The Markup utilise quantitative statistical analysis in addition to investigative reporting, interviews, and document analysis. The Markup's investigation of Amazon's product ranking system, for example, involved training a model to predict where products would appear based on various factors \citep{jeffries_amazon_2021}.  Compared to other domains, these outlets engage in an  
extensive amount of data collection using a variety of custom scraping, data donation, and analysis tools, often built in-house. The Markup's Citizen Browser project, for example, used data donations to investigate discriminatory targeted advertising \citep{angwin_facebook_2016, keegan_facebook_2021, faife_credit_2021} on Facebook \citep{the_markup_citizen_2022}.

\paragraph{Impact} Of the various domains we have examined, journalistic audits result in the most 
impactful outcomes. Audits carried out both by The Markup and ProPublica have resulted in subsequent  
changes in numerous domains including shifting the audit discourse in academia, altering practices in industry (including big tech corporations such as Facebook, Amazon, and Google) \citep{ng_facebook_2021}, and inspiring activism \citep{yin_citing_2021}. Audits both from both outlets 
have also directly inspired legislative and regulatory action \citep{carollo_markups_2021, lecher_nevada_2021}, including a settlement in which the U.S. Department of Justice required Facebook to stop using a special audience tool for housing ads \citep{benner_facebook_2019, feiner_doj_2022}. Those actions included abolishing deployed tools---such as an algorithm governing liver transplants that favored rich, urban patients \citep{carollo_algorithm_2023, carollo_poorer_2023}. ProPublica’s audit of COMPAS set the precedent for algorithmic auditing and remains a canonical work not only for academic research but also as a prime example of algorithmic audit \citep{angwin_machine_2016}. 

\subsection{\textbf{Civil society}}

Activist and other civil society organisations---such as the Electronic Privacy Information Center (EPIC), Data \& Society, and the AI Now Institute---conduct algorithm audits studies as part of their work. We examined 
work from six prominent organisations: 
the Electronic Frontier Foundation (EFF)~\citep{electronic_frontier_foundation_about_2007}, Refugee Law Lab (RLL)~\citep{refugee_law_lab_refugee_2020}, The Citizen Lab~\citep{the_citizen_lab_about_nodate}, Migration Tech Monitor (MTM)~\citep{migration_tech_monitor_migration_2023}, the Ada Lovelace Institute~\citep{ada_lovelace_institute_about_2023} and the American Civil Liberties Union (ACLU)~\citep{american_civil_liberties_union_home_nodate}.

\paragraph{Context} These organisations and institutions primarily conduct case studies, while the Ada Lovelace Institute in particular, also conducts meta-commentary work, especially on audit methods. Though the audits are consistently ~\emph{external}, the harms and concerns these civil society audits aim to surface, diagnose, and mitigate vary. 
Diagnosing and mitigating security vulnerabilities, spying technology, and illegal surveillance are some of the main focuses for EFF. Similarly, the Citizen Lab audits are driven by overarching goals such as investigating and exposing security vulnerabilities and defending free speech online \citep{the_citizen_lab_about_nodate}. RLL and MTM primarily focus on investigative and advocacy work around the questions of transparency and the impacts of technology on refugees \citep{migration_tech_monitor_about_nodate, refugee_law_lab_refugee_2020}. MTM particularly aims to document, map, and monitor migration technology and dismantle and destabilise hierarchical power structures \citep{migration_tech_monitor_about_nodate}. Audits at Ada Lovelace Institute are often driven 
by the objective of policy change with the primary focus of 
diagnosis and remedy for concerns such as privacy, transparency, participation, and disparate impact. Accountability through legal action, litigation, and legal objectives are central to ACLU. 

The audits we examined target mostly ADS in government services, online platforms, and---more than any other domain---surveillance tech. RLL in particular targets technology such as lie detectors and border patrolling drones \citep{refugee_law_lab_refugee_2020}. The Citizen Lab focuses on surveillance and biometric technologies (e.g. for facial recognition), hardware (such as phones and other devices used by politicians and activists), apps, and code \citep{the_citizen_lab_about_nodate}. Biometric data, digital ID systems, surveillance drones, facial recognition, iris scan data, algorithmic motion detectors, ankle monitors, GPS tags, AI powered satellites as well as border surveillance vendors themselves are the central audit targets for MTM \citep{migration_tech_monitor_migration_2023}. The Ada Lovelace Institute also has a diverse target of audit, with a general focus on biometric data and healthcare in particular \citep{ada_lovelace_institute_about_2023}. The ACLU similarly targets facial recognition technology, medical algorithms, welfare algorithms, insurance pricing algorithms, and redlining algorithms as well as advertising algorithms on platforms such as Facebook \citep{american_civil_liberties_union_home_nodate}.

\paragraph{Methodology} Civil society audits utilize the 
most diverse audit methods. EFF, for example, uses policy analysis, grass-root activism, and technology development. Some of the methods used by RLL include interviews with refugees, film making, data collection (from the Immigration and refugee board through Access to Information request, for instance) and analysis, interactive visualisation of data, and documentation of issues such as deportation and refused refugee claims. The Citizen Lab audit methods include in-house developed tools, interviews, and forensic analysis of devices. Similar to RLL, MTM also uses methods such as interviews with people crossing borders, photography, investigative analysis, as well as documenting and archiving migration tech. The main audit methods for Ada Lovelace Institute are participatory methods, including opinion polling, as well as policy research and development. The ACLU often uses quantitative evaluation of, for example operational systems, analysis of data acquired through privileged access, public record requests, and scraping. 

\paragraph{Impact} Similar to journalistic investigations,  
civil 
society audits often result in 
significant impact, often through legal action. These include nuanced and difficult to measure impacts such as shifting public attitude towards surveillance technology or drawing public and media attention towards harmful or controversial tech, as well as 
more concrete outcomes such as moratoriums and abolition. The EFF, for example, won a student's civil lawsuit against the exam surveillance company Proctorio~\citep{gullo_eff_2022}.
In \emph{K.W. v. Armstrong}, the ACLU obtained an injunction stopping algorithmically-determined welfare cuts targeting individuals with developmental disabilities in Idaho \citep{aclu_of_idaho_kw_2016}.
Civil society organisations often represent those harmed by AI systems and in doing so, shift power from the most to the least powerful. 

\subsection{\textbf{Government}}
Several government agencies, especially in Europe, have begun to engage in AI audits. The EU's recent 
Digital Services Act and an AI hiring bill passed in New York City \citep{cumbo_local_2021}, for example, both require some form independent audit. We looked at two government organizations, one with an initially prominent role in AI auditing in the UK, the Information Commissioner's Office (ICO)~\citep{information_commissioners_office_information_2023}, and another in the U.S.,  the National Institute for Standards and Technology (NIST)~\citep{national_institute_of_standards_and_technology_national_2023}.

\paragraph{Context} ICO conducts audits with the broader objective of upholding information rights and enforcing legal requirements \citep{information_commissioners_office_ico_2023, information_commissioners_office_audits_2022}. NIST, on the other hand, is primarily focused on establishing standards and best practice principles for ensuring the development and deployment of socially responsible algorithmic systems \citep{national_institute_of_standards_and_technology_airc_team_nist_nodate, information_commissioners_office_audits_2022}. ICO conducts audits on a case by case basis, 
often published as a report. Comparatively,  NIST 
tends to produce meta-commentary on audits. The central goal for engaging in audit practices for ICO is primarily to investigate and enforce regulations, while NIST lacks such authoritative power and is mainly focused on establishing standards. 

The type of audits regulators carry out tend to be procedural rather than substantive. Audits are carried out with specific objectives which often involve ensuring that a given organisation, institution or corporation is in compliance with established standards or legal requirements. While some organisations like ICO, for example, have the authority to assess and enforce regulatory compliance, others, organisations like NIST, focus on establishing best practice standards and principles that tech developers and vendors are only encouraged to follow. Guidelines from both ICO~\cite{information_commissioners_office_audits_2022} and NIST~\cite{ai2023artificial} tend to apply to informing and standardizing practice amongst ~\emph{internal} audit actors. As these guidelines explicitly mention model development interventions such as fairness or explainability mitigation strategies, it is implied that they apply to ex ante and in media res audits, as well as ex post audits. 

The main audit targets for ICO are data governance practices within public and private companies, public authorities, and government departments, especially those considered to have major impacts \citep{information_commissioners_office_annex_2023, information_commissioners_office_audits_2022}. Organisation can request to be audited but also the ICO audits organisations that the Commissioner believes require audits, that is organisations and corporations with major impacts on society. The ICO audits are a mechanism to check for compliance with data protection legislation, how personal data is managed, data sharing agreements with third parties, and  
security measures, amongst other things. NIST, on the other hand, focuses on artefacts such as facial recognition technology and “general” AI as the main target for audit with the aim of assessing harms such as racial disparities, basic functionality, privacy harms, as well as physical safety of these systems~\cite{grother_face_2019,phillips2003face}. To the degree that government actors  
participate in the audit process themselves, they tend to execute ~\emph{external} ex-post audits, separate from formal corporate participation. 

\paragraph{Methodology} The ICO has developed a set of definitions of what an audit consists of as well as setting out mechanisms for assessing compliance in accordance with regulations based on four assurance ratings: high, reasonable, limited, and very limited assurance. Following an audit, ICO makes recommendations on how to improve on the audit result. The main method used to conduct audits for ICO include examining documents, testing, and interviews with key personnel \citep{information_commissioners_office_audits_2022}. For NIST, the main audit methods 
include data trust, accuracy evaluation, and benchmarking \citep{national_institute_of_standards_and_technology_airc_team_nist_nodate, ngan_ongoing_2020}. 

\paragraph{Impact}
ICO has issued several monetary penalties for data protection legislation breaches, including a fine of £12.7 million to TikTok for misusing children’s data~\citep{information_commissioners_office_ico_2023} and £17 million fine to Clearview AI inc~\citep{information_commissioners_office_ico_2021}, 
in addition to issuing a notice to stop further processing of the personal data of people in the UK and a request to delete such data. NIST has made significant impact in the U.S. regulatory conversation with its AI Risk Management Framework (RMF), though it has not conducted any audits with specific impacts and has no power to enforce these guidelines \citep{tabassi_ai_2023}.

\section{Discussion}

While many of the academic studies we reviewed formed the foundation for important methods and topics for AI auditing, they also often lacked in achieving comparable levels of impact to the audit work we analyzed in journalism, civil society, government, and industry. 
In this section, we outline several practical takeaways from our analysis for researchers, policymakers, and practitioners. In particular, we identify several ways that audit work outside academia could serve as a guide for more impactful audit studies within academia: 
first, by expanding the aims of audit work beyond evaluation, and towards accountability; second, by encouraging more explicit and forward-looking engagement with often excluded stakeholders, acknowledging power asymmetries~\citep{kalluri2020don}, and fostering mechanisms 
for collective action and ecological change; and third, by offering specific improvements for audit methodology, including specificity, practitioner diversity, and interdisciplinary collaboration.

\subsubsection{\textbf{Power asymmetries \& auditor-stakeholder relationships}} 

One of the most influential factors in determining an audit's impact was revealed in the power analysis of how the auditors interacted with various stakeholders---including impacted populations, the audit target and others. 
Subsequently, methodological innovation and evaluation of AI systems without considerations of structural factors---such as the uneven distribution of power, control, benefit and harm---is unlikely to result in significant impact. 
Future research could further explore tools and strategies to map out, define and foster auditor relationships with various stakeholders. Similarly, policy-makers and other authorities 
could step in to empower auditors -- 
for example, by incorporating audit results into more consequential repercussions for companies or open-sourcing, publishing, disseminating, and reviewing submitted audit results.

\subsubsection{\textbf{Prioritizing audit execution stages beyond evaluation}}
Framing details such as the 
selection of audit targets, named motivations, types of harms investigated and 
the scope of audit goals are more likely to determine the effectiveness of an audit's outcome than the details of audit execution. Most of the academic work we reviewed focused on the process of evaluating AI systems for bias, fairness, or disparate impacts. Conversely, these studies rarely focused on other stages of auditing crucial to accountability in non-academic work, such as discovering harms, communicating audit results, or organizing non-technical interventions and collective action. As detailed in Section~\ref{sec:other-domains}, 
these factors are critical in influencing audit success outside of the auditor's control ~\citep{raji_outsider_2022, costanza-chock_who_2022}. For example, some auditors, such as regulators and law firms, have the legal authority to demand access. 
Meanwhile, for others, this degree of internal visibility is rare, requiring the auditors to rely on proxy evaluations and open up their audit results to critique, denial and retaliation from the audit targets. 

Even as policymakers work to install legal protections for external auditors and direct resources towards harms discovery and audit reporting, future research should investigate the under-studied aspects of the AI auditing process and practitioners should prioritize these aspects of auditing in addition to evaluation.

\subsubsection{
\textbf{Expanding audit scope beyond the product, model or algorithm}} 
In our survey, we use the ecosystem audit classification to describe studies that consider the entire AI pipeline in a holistic way--- communities and socio-technical environments (digital or physical) defining or impacted by critical components to an AI system's operation---
(\S\ref{sec:case-studies}). While these ecosystem audits exemplify many of the elements of the socio-technical auditing advocated for in various academic critiques \citep{gansky_counterfacctual_2022, radiya-dixit_sociotechnical_2023, lam_sociotechnical_2023}, most academic audit studies we found 
extremely rarely involved a comprehensive analysis of involved stakeholders and any holistic view of multiple interacting 
AI systems.  
Future work should continue to incorporate and expand broader and holistic perspectives in order to craft a richer account of algorithmic systems and their impacts.

\subsubsection{\textbf{Impact increases with specificity}}
Even as audit studies broaden their scope to include the ecosystem surrounding and defining an AI system, most audits need to be specific in order to be more impactful. As has been previously discussed in the aftermath of the Gender Shades audit ~\citet{raji_actionable_2022}, naming a specific audit target, advocacy objectives and intended target responses ensures that the audit speaks to more specific demands from the audit target. This makes that target more likely to 
to respond to the audit and informs advocates on precise policy demands. This strategy was seen in many civic audits, such as the ACLU, the Markup and regulators like ICO, where 
the scope and demands tied to their investigations are specific. 
When audit scope was too diffused, the audit seemed to hold less effectiveness for accountability. ORCAA's Hirevue audit, for example, was criticized for being too high level to inform any specific allegations or demands for remedy~\cite{engler_independent_2021}.

Constraining audit objectives and scope has benefits as well as limitations. Too narrow a scope may obfuscate broader systematic factors and make it even more difficult to advocate for more structural changes beyond the scope of the examined audit target~\citep{raji2020saving, raji_actionable_2022}. Therefore, future work should aim to be highly specific without losing focus on holistic and ecological observations.

\subsubsection{\textbf{Utilising 
a wider range of audit methodologies}}

Within CS and CS-adjacent academic venues, quantitative methodological approaches to AI auditing were often front and center. These methods are evident in the range of fairness metrics developed and debated within these communities~\cite{friedler2021possibility}. Existing critiques already articulate the limitations of these approaches, in terms of its abstraction~\cite{selbst_fairness_2019}, legal incompatibilities~\cite{xiang_reconciling_2020, xiang_legal_2019} and lack of connection to substantive outcomes~\cite{green2022escaping}. 
It is particularly noteworthy that in many non-academic domains, the methods employed are far from what is being explored or commonly discussed in academia, including a range of qualitative methodological approaches such as investigative reporting, document review, and stakeholder consultation. Granted, these other institutions often possess different skill sets and operate within different constraints. However, especially for academics hoping to connect their work to the practice of other audit practitioner communities, future AI audit research should explore how extra-disciplinary methods and tactics like these---especially those with a proven record of impact---could bolster academic audit work. 

\subsubsection{\textbf{Appreciating the diversity of audit practitioners}}

Different audit communities possess different strengths and weaknesses---civil society auditors, for example, typically have mature and well-developed methods around audit communication and advocacy, but at times may lack the technical capacity to execute more complex quantitative analyses.  
On the other hand, in academia, auditors 
might be well-equipped for innovating technically but struggled with public and other stakeholder engagement.
 
Our survey supports prior findings~\cite{costanza-chock_who_2022} that audit practitioners within a given domain are not a monolith. For instance, there are law firms that provide audits as a service 
on behalf of audit targets while others execute civic audits on behalf of, and in collaboration with impacted populations. Similarly, we found 
a wide range of technical competence within 
domains. 
Audits from the journalism organization the Markup, for instance, were often more thorough and reproducible than some academic studies. Similarly, auditors can have differing motivations and goals, which deeply informs their practice~\cite{laufer_strategic_2023}. 

Future research on the practice of AI auditing should still consider the strengths and weaknesses of different auditing institutions. These findings are informative 
not only for improving audit practice but also 1) facilitating 
avenues for collaboration with auditors outside academia and 2) informing a growing set of enacted and proposed legislation requiring audits for AI systems \citep{cumbo_local_2021, clarke_algorithmic_2019, lenhart_federal_2023, perrigo_california_2023, mendelson_stop_2021}.

\subsubsection{\textbf{Timing 
and auditor type are not the primary determining factors of audit impact}}

The distinction of~\emph{when} the audit should ideally be conducted (i.e. ex ante, in media res, ex post) has recently been the subject of some policy debate~\cite{zero_trust_ai_now, stein_safe_2023}. As expected, those with increased access (i.e. internal auditors) typically have the opportunity to execute audits pre-deployment, while others (i.e. external auditors) are often restricted to conducting audits post-deployment. However, when it comes to translating audit results to accountability, 
both scenarios come with 
distinct challenges and it seems that other contextual factors outside of auditor type or timing, including audit goal, target, design, communication and scope, can ultimately be much  
more meaningful. 

For example, despite increased and early access, internal auditors may still struggle to convince key organizational stakeholders to act on audit results~\cite{phan2022economies,boag2022tech,widder2023s}. Furthermore, such auditors can become 
vulnerable to internal corporate retaliation, and corporate censorship~\cite{ryan2022ai} as well as being mired in 
conflict of interest~\cite{schellmann2021auditors}. All of this can interfere with the auditors' ability to externally communicate results, or force auditors to scope results too narrowly to be meaningfully.  
In policy, internal auditors are typically those designated to carry out mandatory corporate audit requirements (e.g. the ``independent auditors'' in Article 37 of the Digital Services Act). The particular challenges of internal auditors indicate that requirements for external reporting, product standards and auditor conduct standards are especially important in such contexts.

On the other hand, external auditors are free to set their scope and communicate results publicly. However, they typically face issues around external legitimacy and visibility that can make them easy to ignore completely~\cite{raji_actionable_2022}. The serious issues they face around auditor capacity and information access are further hindrances to audit quality and thus impact~\cite{raji_outsider_2022}, and the public-facing nature of their advocacy makes them even more vulnerable to corporate retaliation. External auditors are typically designated as voluntary ``investigators'' or ``researchers'', rather than actual auditors  (e.g. ``vetted researchers'' in Article 40 of the Digital Services Act), and are often mentioned in data access or safe harbor clauses, revealing the importance of such policy interventions in addressing their particular challenges. 

\subsubsection{\textbf{Recognizing the limits of audits}} 

Some societal impacts of AI are not amenable to audits. What can be evaluated and audited depends on numerous factors including normative and pragmatic considerations as well as what is prioritised and anticipated, and whether harms and risks are known or anticipated~\citep{weidinger_sociotechnical_2023}. As there are several serious, although  gradual, and pernicious harms emanating from AI -- for instance, the chilling effect of surveillance, or corporate power concentration~\citep{birhane2022values} -- it is clear that not all pressing issues regarding the technology can be addressed with audits. Although the ecological audits we have introduced in this work are an encouraging first step towards significant structural change in some of these cases, other, likely moral or rights-based arguments are required for this kind of advocacy.
Thus, it is clear that audits are just one element of a necessarily broader set of AI accountability strategies. We caution against any approaches that treat audits as any kind of all-encompassing solution to technological ills. 

\section{Conclusion}

Audits have become widely popular in the AI field as  an aspirational accountability measure to inform our decision-making and regulation of AI deployments. However, lessons from the broader range of AI audit practitioners beyond academia reveal that there is still much to learn in order for these audit studies to 
operate as consequential judgements. Future work will hopefully continue to investigate this relationship between contextual factors, audit design, audit execution and the underlying shared vision of audit practitioners across all domains: a substantively meaningful path towards AI accountability.

\bibliographystyle{IEEEtranN}
\bibliography{IEEEabrv,ryan-biblatex.bib,additional.bib}

\appendices
\renewcommand\thetable{\thesection-\Roman{table}}
\setcounter{table}{0}
\renewcommand\thefigure{\thesection-\Roman{figure}}
\setcounter{figure}{0}

\section{Methods: Searching in Non-Academic Domains}
\label{app:supplementary}
To produce the analysis in \S\ref{sec:other-domains}, we analyzed reports, press releases, publications, blogs, and web postings from multiple institutions outside of academia. Here, we describe in more detail the specific search methods used to produce our analysis for each institution. (The full list of academic audit studies we found can be accessed at {\github}.)

\subsection{Regulators}
\subsubsection{National Institute of Standards and Technology (NIST)} We reviewed documents ($N=16$) from two programs: NIST's Trustworthy \& Responsible AI Resource Center (\href{https://airc.nist.gov/}{airc.nist.gov}) and NIST's Face Recognition Vendor Test (FRVT) program \citep{national_institute_of_standards_and_technology_face_2020}. From the Resource Center, we reviewed the NIST AI Risk Management Framework (RMF) \citep{tabassi_ai_2023} (as well as the accompanying Playbook and Roadmap) and the $N=5$ special publications and internal reports in the RMF Knowledge Base \citep{national_institute_of_standards_and_technology_airc_team_nist_nodate}. From the FRVT program, we reviewed the executive summary of all the pages linked under sections with titles pre-pended ``FRVT'' ($N=8$) on the FRVT page \citep{national_institute_of_standards_and_technology_face_2000, national_institute_of_standards_and_technology_face_2002, national_institute_of_standards_and_technology_face_2006, national_institute_of_standards_and_technology_face_2013, national_institute_of_standards_and_technology_face_2020}, including the homepages for FRVT 2000, 2002, 2006, and 2013, as well as NIST Interagency Reports (NISTIR) 8331, 8280, 7709, and 7830. NISTIR 8280 in particular describes analyses of demographic differences across a large number of algorithms, developers, and test images \citep{grother_face_2019}.

\subsubsection{Information Commissioner's Office (ICO)}
The ICO details audit information on its website \href{https://ico.org.uk/}{ico.org.uk}, including definition of audit, assurance ratings, audit target, as well as audit reports. We reviewed the \href{https://ico.org.uk/action-weve-taken/audits-and-overview-reports/}{``Audits and overview reports''} section yielding 48 results in total. 

\subsection{Law Firms}
\subsubsection{AWO} We reviewed publicly available pages on AWO's website (\href{https://www.awo.agency}{awo.agency}), particularly the descriptions of their services at \href{https://www.awo.agency/services}{awo.agency/services}. We also reviewed all the posts on their blog up to July 2023 ($N=62$), only a few of which were related to AWO's audit work (\citet{awo_announcing_2023}, for example).
\subsubsection{Foxglove}
In order to extract relevant information on the organization, we reviewed the website, mainly the~\href{https://www.foxglove.org.uk/who-we-are/}{Who We Are} and~\href{https://www.foxglove.org.uk/news/}{News} sections. 
\subsubsection{Luminos.Law}
Our information on Luminos.Law---called BHN.AI at the time of our study---is sourced from the~\href{https://luminos.law/resources}{firm's website}, and more particularly, in the~\href{https://luminos.law/ai-audits}{AI Audits section}.  

\subsection{Civil Society}
\subsubsection{Electronic Frontier Foundation (EFF)}
We looked at the~\href{https://www.eff.org/about}{About} page to extract general information on the Foundation. We also reviewed the~\href{https://www.eff.org/updates}{Our Work}, and~\href{https://www.eff.org/updates}{Tools} page in order to extract detailed information on including Methods and Tools. 
\subsubsection{Refugee Law Laboratory (RLL)}
The Refugee Law Lab website~\href{https://refugeelab.ca/}{About} page provided the initial general information. For reports, datasets, data visualizations, and impact, we reviewed the~\href{https://refugeelab.ca/projects/refugee-law-data/}{Projects} and~\href{https://refugeelab.ca/publications/}{News and Recent Publications} sections. 
\subsubsection{The Citizen Lab}
The~\href{https://citizenlab.ca/about/}{About the Citizen Lab} provided general information about this group. In order to extract detailed relevant information, we reviewed the~\href{https://citizenlab.ca/category/research/}{Research} and~\href{https://citizenlab.ca/category/lab-news/}{News} pages. 
\subsubsection{Migration and Tech Monitor (MTM)}
We started our survey with the~\href{https://www.migrationtechmonitor.com/}{Home} page of MTM's website. The~\href{https://www.migrationtechmonitor.com/about-us}{Methodology} section details the methods used by the organization, the~\href{https://www.migrationtechmonitor.com/resources}{Resources} page lists various case studies carried out by the organization, and the~\href{https://www.migrationtechmonitor.com/snapshots}{Snapshots} section presents numerous pictorial evidence supporting the organization's work.  

\subsubsection{Ada Lovelace Institute} We surveyed the ``Projects'' listed at \href{https://www.adalovelaceinstitute.org/our-work/}{adalovelaceinstitute.org/our-work}, selecting only those related to AI auditing. We followed any reports listed in those project overviews. In particular, we analyzed the executive summary of $N=4$ reports on algorithmic impact assessments in healthcare \citep{groves_algorithmic_2022}, tools and methods for assessing algorithmic systems \citep{ada_lovelace_institute_examining_2020, ada_lovelace_institute_technical_2021}, and auditing standards \citep{brennan_getting_2022}. We also surveyed $N=47$ blog posts published July 2023 or earlier at \href{https://www.adalovelaceinstitute.org/blog}{adalovelaceinstitute.org/blog}, \href{https://www.adalovelaceinstitute.org/blog/?select-programmes%5B%5D=biometrics&select-programmes%5B%5D=enabling-responsible-ai-ecosystem&select-programmes%5B%5D=ethics-and-accountability-in-practice&select-programmes%5B%5D=public-sector-use-of-data-algorithms&hidden-s=&hidden-current-page=1#listing}{filtering} for the programmes ``Biometrics'', ``Enabling a responsible AI ecosystem'', ``Ethics and accountability in practice'', and ``Public-sector use of data \& algorithms''. We considered only blog posts that discussed the Institute's own AI auditing work.

\subsubsection{The American Civil Liberties Union (ACLU)}
The ACLU publishes details about its work on its website \href{https://www.aclu.org}{aclu.org}, including news, publications and reports. We reviewed all documents labelled as ``reports and other assets'' published 2018--2022 ($N=70$), using \href{https://www.aclu.org/search}{aclu.org/search}. Most of these did not relate to AI, so we also used Google to search \href{https://www.aclu.org/news}{aclu.org/news} using the keywords from our academic search as well as other keywords such as ``algorithm'', ``machine learning'', and ``automated decision'' (e.g., \texttt{site:aclu.org/news after:2017 ``algorithm''}).

\subsection{Journalism}
\subsubsection{Propublica}
We initially scraped the~\href{https://www.propublica.org/}{ProPublica} website using the keyword ``audit''. However, this did not yield fruitful as significant number of articles that the keyword returned were not relevant, for example,~\href{https://www.propublica.org/article/gao-slams-flimsy-auditing-rules-for-stimulus-dollars-709}{articles on financial audit} and~\href{https://projects.propublica.org/coronavirus-contracts/contracts/75P00119A00032}{web pages with reports of financial audits}. Subsequently, we manually reviewed the website looking specifically for algorithm (and technology) related audits under the~\href{https://www.propublica.org/topics/technology}{Technology} section of the website. 
\subsubsection{The Markup}
We manually sifted through the main page of the~\href{https://themarkup.org/series/machine-learning}{Markup's website} identifying audit investigation report articles. We reviewed each article and extracted thematic information.

\subsection{Consulting Agencies}
\subsubsection{O'Neil Risk Consulting \& Algorithmic Auditing (ORCAA)}
We reviewed the website starting from the~\href{https://orcaarisk.com/}{Home} page and scrolling through the main pages such as~\href{https://orcaarisk.com/}{What We Do},~\href{https://orcaarisk.com/}{NYC Bias Audit} and~\href{https://orcaarisk.com/}{Principles}.  
\subsubsection{Eticas}
We surveyed the~\href{https://eticas.tech/algorithmic-audits}{Eticas website}, more particularly focusing on the~\href{https://eticas.tech/research}{Research} page of the Eticas Library to glean insights into the organization's algorithmic audit practice. 
\subsubsection{BABL AI}
We started by reviewing the~\href{https://babl.ai/about-us/}{About Us} page of BABL.IA'S website. We then looked at the~\href{https://babl.ai/services/}{Services} page and extracted contextual information such as the kind of information the agency provides. We also reviewed the~\href{https://babl.ai/courses/}{Research} page for more detailed information on past audits carried out by the agency.

\section{Methods: Quantitative Content Analysis}
\label{app:keyword}
We supplement our qualitative analysis in \S\ref{sec:academic} by analyzing words and phrases commonly used in the abstracts and keywords of the academic audit studies we collected. All but {\nMissingAbstract} of the $N={\nAcademic}$ studies we found were published with an abstract; {\nMissingKeywords} were published without keywords.

 Many of the most-used terms are fairly generic (e.g. ``algorithm'', or ``system), so we show only a selection of key terms relevant to our study. We manually filtered the most frequent 1-grams and 2-grams across all abstracts and keywords in our dataset of academic audits, keeping only those terms relevant to one of the following dimensions of our analysis (see Table~\ref{tab:summary}): 1) motivation (e.g., ``accountability''); 2) target (including specific entities, e.g., ``Facebook''; types of systems, e.g., ``computer vision''; and domains, e.g., ``healthcare'' or ``hiring''); 3) types of harms (e.g., ``discrimination'' or ``privacy''); 4) methods (e.g., ``qualitative'' or ``ethnography''). We drop 2-grams that are already encompassed by a more general 1-gram unless they indicate a meaningfully different concept or subset (e.g., ``criminal justice'' is much different than ``justice'' alone, whereas, e.g., ``algorithm auditing'' is encompassed by ``auditing'', given the scope of our study). We include only terms that were mentioned in more than one keyword list (more than 0.27\% of papers) or in more than 10 abstracts (more than 2.7\% of papers).
 
 For further analysis, the entire dataset of academic audits we collected, along with the code for this analysis, can be accessed at {\github}.

 \todo[camera-ready: make the code public, too, so people can see the list of keywords]

Tables~\ref{tab:kwds-motivation}--\ref{tab:kwds-methods} list the most frequent terms related to audit motivations and harms, audit targets, and audit methods, respectively. Tables~\ref{tab:kwds-motivation-pivot}--\ref{tab:kwds-methods-pivot} further compare the most frequent keywords across audit types.

 \begin{table*}[!ht]
\caption{Most frequent terms: audit motivation and types of harms.}
\centering
\label{tab:kwds-motivation}
\begin{tabularx}{0.6\linewidth}{X p{4em} p{5em}}
\toprule
\textbf{Term(s)} & \textbf{\# times used} & \textbf{\# studies (\% of total)} \\
\midrule

\textit{Terms in keywords (only terms in $\geq1\%$ of studies)} & & \\
fairness/fair &  75 & 66 (25.1\%) \\ 
bias/biases/biased &  75 & 65 (24.7\%) \\ 
audit/audits/auditing &  56 & 49 (18.6\%) \\ 
accountability/accountable &  42 & 41 (15.6\%) \\ 
ethics/ethical &  26 & 26 (9.9\%) \\ 
impact/impacts &  17 & 15 (5.7\%) \\ 
evaluate/evaluation/evaluations/evaluating &  17 & 13 (4.9\%) \\ 
privacy &  16 & 12 (4.6\%) \\ 
transparency &  12 & 12 (4.6\%) \\ 
explainability/explanations/explanation/xai &  12 & 10 (3.8\%) \\ 
discrimination/discriminatory &   8 & 8 (3.0\%) \\ 
justice &   8 & 8 (3.0\%) \\ 
governance &   6 & 6 (2.3\%) \\ 
hate speech &   6 & 6 (2.3\%) \\ 
impact assessment(s) &   6 & 6 (2.3\%) \\ 
disparate impact / disparity / disparities &   5 & 5 (1.9\%) \\ 
gender bias &   5 & 5 (1.9\%) \\ 
harm/harms &   5 & 5 (1.9\%) \\ 
risk &   5 & 5 (1.9\%) \\ 
data protection &   5 & 4 (1.5\%) \\ 
interpretability &   4 & 4 (1.5\%) \\ 
misinformation &   6 & 4 (1.5\%) \\ 
security &   4 & 4 (1.5\%) \\ 
surveillance &   4 & 4 (1.5\%) \\ 
trust &   4 & 4 (1.5\%) \\ 
accessibility &   3 & 3 (1.1\%) \\ 
\midrule

\textit{Terms in abstract (only terms in $\geq5\%$ of studies)} & & \\
bias/biases/biased & 325 & 99 (29.4\%) \\ 
audit/audits/auditing & 201 & 87 (25.8\%) \\ 
impact/impacts & 135 & 85 (25.2\%) \\ 
fairness/fair & 271 & 78 (23.1\%) \\ 
evaluate/evaluation/evaluations/evaluating & 114 & 69 (20.5\%) \\ 
accountability/accountable & 112 & 56 (16.6\%) \\ 
harm/harms &  74 & 44 (13.1\%) \\ 
transparency &  62 & 44 (13.1\%) \\ 
ethics/ethical &  65 & 37 (11.0\%) \\ 
discrimination/discriminatory &  57 & 34 (10.1\%) \\ 
risk &  55 & 34 (10.1\%) \\ 
mitigate &  31 & 28 (8.3\%) \\ 
privacy &  56 & 25 (7.4\%) \\ 
explainability/explanations/explanation/xai &  63 & 22 (6.5\%) \\ 
disparate impact / disparity / disparities &  37 & 21 (6.2\%) \\ 
justice &  31 & 21 (6.2\%) \\ 
\bottomrule
\end{tabularx}
\end{table*}
 \begin{table*}[!ht]
\caption{Most frequent terms: target systems and domains.}
\centering
\label{tab:kwds-targets}
\begin{tabularx}{0.6\linewidth}{X p{4em} p{5em}}
\toprule
\textbf{Term(s)} & \textbf{\# times used} & \textbf{\# studies (\% of total)} \\
\midrule

\textit{Terms in keywords (only terms in $\geq1\%$ of studies)} & & \\
social media / social networks / facebook / twitter / youtube &  18 & 17 (6.5\%) \\ 
  computer vision &  13 & 13 (4.9\%) \\ 
  health/healthcare/medical &  14 & 11 (4.2\%) \\ 
  natural language / language processing &  22 & 11 (4.2\%) \\ 
  advertising/ads &   8 & 7 (2.7\%) \\ 
  platform/platforms &   6 & 6 (2.3\%) \\ 
  hiring/employment &   5 & 5 (1.9\%) \\ 
  policing/police &   5 & 5 (1.9\%) \\ 
  credit &   4 & 4 (1.5\%) \\ 
  criminal justice &   4 & 4 (1.5\%) \\ 
  multimodal &   4 & 4 (1.5\%) \\ 
  search engines &   4 & 4 (1.5\%) \\ 
  speech recognition / speaker recognition / speaker verification &   6 & 4 (1.5\%) \\ 
  automated decision &   3 & 3 (1.1\%) \\ 
  facial recognition / face recognition &   3 & 3 (1.1\%) \\ 
  generative &   3 & 3 (1.1\%) \\ 
  google &   3 & 3 (1.1\%) \\ 
  government &   3 & 3 (1.1\%) \\ 
  risk assessment &   3 & 3 (1.1\%) \\ 
  social credit &   3 & 3 (1.1\%) \\ 
  welfare &   3 & 3 (1.1\%) \\

\midrule

\textit{Terms in abstract (only terms in $\geq5\%$ of studies)} & & \\
platform/platforms &  99 & 45 (13.4\%) \\ 
  health/healthcare/medical &  89 & 42 (12.5\%) \\ 
  social media / social networks / facebook / twitter / youtube &  99 & 40 (11.9\%) \\ 
  advertising/ads &  79 & 21 (6.2\%) \\ 
  hiring/employment &  23 & 18 (5.3\%) \\ 
  google &  22 & 17 (5.0\%) \\ 
  government &  22 & 17 (5.0\%) \\ 
\bottomrule
\end{tabularx}
\end{table*}
 \begin{table*}[!ht]
\caption{Most frequent terms: methods.}
\centering
\label{tab:kwds-methods}
\begin{tabularx}{0.6\linewidth}{X p{4em} p{5em}}
\toprule
\textbf{Term(s)} & \textbf{\# times used} & \textbf{\# studies (\% of total)} \\
\midrule

\textit{Terms in keywords (only terms in $\geq1\%$ of studies)} & & \\
sociotechnical &   9 & 9 (3.4\%) \\ 
  participatory &   8 & 7 (2.7\%) \\ 
  hci &   7 & 6 (2.3\%) \\ 
  qualitative/ethnography/interviews/interviewed/workshop(s) &   7 & 5 (1.9\%) \\ 
  participatory design &   5 & 5 (1.9\%) \\ 
  community/communities &   4 & 4 (1.5\%) \\ 
  human centered &   4 & 4 (1.5\%) \\ 
  interdisciplinary &   5 & 3 (1.1\%) \\

\midrule

\textit{Terms in abstract (only terms in $\geq5\%$ of studies)} & & \\
community/communities &  70 & 48 (14.2\%) \\ 
  qualitative/ethnography/interviews/interviewed/workshop(s) &  37 & 26 (7.7\%) \\ 
  benchmark/benchmarks &  50 & 24 (7.1\%) \\ 
  sociotechnical &  27 & 22 (6.5\%) \\ 

\bottomrule
\end{tabularx}
\end{table*}
 \begin{table*}[!ht]
\caption{Frequent terms by audit type: audit motivation and types of harms.}
\centering
\label{tab:kwds-motivation-pivot}
\begin{tabularx}{0.8\linewidth}{X p{6em} p{6em} p{6em} p{6em}}
\toprule
\multirow{2}{*}{\textbf{Term(s)}} & \multicolumn{4}{c}{\textbf{\# studies (\% of studies of the same audit type)}} \\
& \emph{Data Audit} & \emph{Product/Model/ Algo. Audit} & \emph{Ecosystem Audit} & \emph{Meta-Commentary} \\
\midrule

\textit{Terms in keywords} & & & & \\
bias/biases/biased & 8 (\textbf{29.6}\%) & 48 (\textbf{28.4}\%) & 3 (\textbf{20.0}\%) & 6 (11.5\%) \\ 
  fairness/fair & 8 (\textbf{29.6}\%) & 40 (\textbf{23.7}\%) & 3 (\textbf{20.0}\%) & 15 (\textbf{28.8}\%) \\ 
  audit/audits/auditing & 1 (3.7\%) & 35 (\textbf{20.7}\%) & 1 (6.7\%) & 12 (\textbf{23.1}\%) \\ 
  accountability/accountable & 1 (3.7\%) & 16 (9.5\%) & 3 (\textbf{20.0}\%) & 21 (\textbf{40.4}\%) \\ 
  ethics/ethical & 3 (11.1\%) & 13 (7.7\%) & 0 & 10 (\textbf{19.2}\%) \\  
\midrule
\textit{Terms in abstract} & & & & \\
bias/biases/biased & 18 (\textbf{39.1}\%) & 69 (\textbf{32.5}\%) & 2 (13.3\%) & 10 (\textbf{15.6}\%) \\ 
  fairness/fair & 8 (\textbf{17.4}\%) & 45 (\textbf{21.2}\%) & 4 (\textbf{26.7}\%) & 21 (\textbf{32.8}\%) \\ 
  audit/audits/auditing & 4 (8.7\%) & 57 (\textbf{26.9}\%) & 1 (6.7\%) & 25 (\textbf{39.1}\%) \\ 
  evaluate/evaluation/evaluations/evaluating & 11 (\textbf{23.9}\%) & 45 (\textbf{21.2}\%) & 2 (13.3\%) & 11 (\textbf{17.2}\%) \\ 
  impact/impacts & 12 (\textbf{26.1}\%) & 52 (\textbf{24.5}\%) & 4 (\textbf{26.7}\%) & 17 (\textbf{26.6}\%) \\ 
  accountability/accountable & 4 (8.7\%) & 29 (13.7\%) & 5 (\textbf{33.3}\%) & 18 (\textbf{28.1}\%) \\ 
  transparency & 5 (10.9\%) & 26 (12.3\%) & 4 (\textbf{26.7}\%) & 9 (14.1\%) \\ 
  ethics/ethical & 7 (\textbf{15.2}\%) & 19 (9.0\%) & 2 (13.3\%) & 9 (14.1\%) \\ 
  harm/harms & 8 (\textbf{17.4}\%) & 19 (9.0\%) & 4 (\textbf{26.7}\%) & 13 (\textbf{20.3}\%) \\ 
  risk & 2 (4.3\%) & 20 (9.4\%) & 3 (\textbf{20.0}\%) & 9 (14.1\%) \\ 
  mitigate & 3 (6.5\%) & 18 (8.5\%) & 3 (\textbf{20.0}\%) & 4 (6.2\%) \\ 
\bottomrule
\multicolumn{5}{p{40em}}{\footnotesize Showing only terms that appear in $\geq 15\%$ of studies for at least one audit type (bolded).}
\end{tabularx}
\end{table*}
 \begin{table*}[!ht]
\caption{Frequent terms by audit type: target systems and domains.}
\centering
\label{tab:kwds-targets-pivot}
\begin{tabularx}{0.8\linewidth}{X p{6em} p{6em} p{6em} p{6em}}
\toprule
\multirow{2}{*}{\textbf{Term(s)}} & \multicolumn{4}{c}{\textbf{\# studies (\% of studies of the same audit type)}} \\
& \emph{Data Audit} & \emph{Product/Model/ Algo. Audit} & \emph{Ecosystem Audit} & \emph{Meta-Commentary} \\
\midrule

\textit{Terms in keywords} & & & & \\
social media / social networks / facebook / twitter / youtube & 0 & 17 (\textbf{10.1}\%) & 0 & 0 \\ 
  natural language / language processing & 3 (\textbf{11.1}\%) & 6 (3.6\%) & 1 (6.7\%) & 1 (1.9\%) \\ 
  computer vision & 7 (\textbf{25.9}\%) & 5 (3.0\%) & 0 & 1 (1.9\%) \\ 
  welfare & 1 (3.7\%) & 0 & 2 (\textbf{13.3}\%) & 0 \\ 
  child welfare & 0 & 0 & 2 (\textbf{13.3}\%) & 0 \\ 
\midrule
\textit{Terms in abstract} & & & & \\
social media / social networks / facebook / twitter / youtube & 2 (4.3\%) & 38 (\textbf{17.9}\%) & 0 & 0 \\ 
  platform/platforms & 1 (2.2\%) & 42 (\textbf{19.8}\%) & 1 (6.7\%) & 1 (1.6\%) \\ 
  health/healthcare/medical & 5 (\textbf{10.9}\%) & 30 (\textbf{14.2}\%) & 1 (6.7\%) & 6 (9.4\%) \\ 
  computer vision & 7 (\textbf{15.2}\%) & 8 (3.8\%) & 0 & 1 (1.6\%) \\ 
  natural language / language processing & 5 (\textbf{10.9}\%) & 9 (4.2\%) & 0 & 1 (1.6\%) \\ 
  hiring/employment & 2 (4.3\%) & 11 (5.2\%) & 2 (\textbf{13.3}\%) & 3 (4.7\%) \\ 
  government & 2 (4.3\%) & 8 (3.8\%) & 3 (\textbf{20.0}\%) & 4 (6.2\%) \\ 
  welfare & 1 (2.2\%) & 3 (1.4\%) & 2 (\textbf{13.3}\%) & 0 \\ 
  child welfare & 0 & 1 (0.5\%) & 2 (\textbf{13.3}\%) & 0 \\ 
  education & 1 (2.2\%) & 6 (2.8\%) & 2 (\textbf{13.3}\%) & 1 (1.6\%) \\  
\bottomrule
\multicolumn{5}{p{40em}}{\footnotesize Showing only terms that appear in $\geq 10\%$ of studies for at least one audit type (bolded).}
\end{tabularx}
\end{table*}
 \begin{table*}[!ht]
\caption{Frequent terms by audit type: methods.}
\centering
\label{tab:kwds-methods-pivot}
\begin{tabularx}{0.8\linewidth}{X p{6em} p{6em} p{6em} p{6em}}
\toprule
\multirow{2}{*}{\textbf{Term(s)}} & \multicolumn{4}{c}{\textbf{\# studies (\% of studies of the same audit type)}} \\
& \emph{Data Audit} & \emph{Product/Model/ Algo. Audit} & \emph{Ecosystem Audit} & \emph{Meta-Commentary} \\
\midrule

\textit{Terms in keywords} & & & & \\
qualitative/ethnography/interviews/interviewed/workshop(s) &  & 3 (1.8\%) & 2 (\textbf{13.3}\%) & 0 \\ 
  participatory &  & 1 (0.6\%) & 3 (\textbf{20.0}\%) & 3 (5.8\%) \\ 
  participatory design &  & 0 & 3 (\textbf{20.0}\%) & 2 (3.8\%) \\ 
  human centered &  & 2 (1.2\%) & 2 (\textbf{13.3}\%) & 0 \\   
\midrule
\textit{Terms in abstract} & & & & \\
benchmark/benchmarks & 14 (\textbf{30.4}\%) & 6 (2.8\%) & 0 & 4 (6.2\%) \\ 
  community/communities & 9 (\textbf{19.6}\%) & 21 (9.9\%) & 6 (\textbf{40.0}\%) & 12 (\textbf{18.8}\%) \\ 
  qualitative/ethnography/interviews/interviewed/ workshop(s) & 3 (6.5\%) & 13 (6.1\%) & 6 (\textbf{40.0}\%) & 4 (6.2\%) \\ 
  sociotechnical & 1 (2.2\%) & 12 (5.7\%) & 1 (6.7\%) & 8 (\textbf{12.5}\%) \\ 
  hci & 0 & 3 (1.4\%) & 2 (\textbf{13.3}\%) & 1 (1.6\%) \\ 
  participatory & 1 (2.2\%) & 0 & 2 (\textbf{13.3}\%) & 1 (1.6\%) \\    
\bottomrule
\multicolumn{5}{p{40em}}{\footnotesize Showing only terms that appear in $\geq 10\%$ of studies for at least one audit type (bolded).}
\end{tabularx}
\end{table*}

\section{Additional Results}
Below, we include more detailed summaries of the non-academic auditors we reviewed.
\begin{figure*}[!t]
    \centerline{\includegraphics[width=0.5\linewidth]{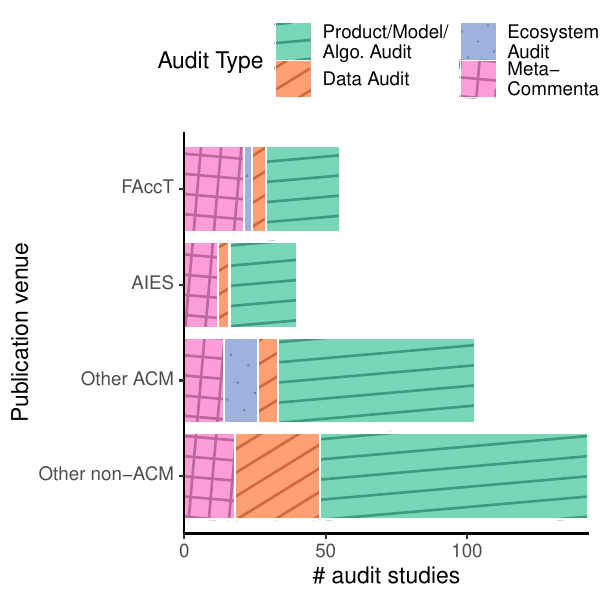}}
    \caption{Number of collected academic audit studies published in each year, grouped by publication source.}
    \label{fig:bar-source}
\end{figure*}

\begin{table*}
\centering
\caption{Consulting Agencies}
\label{tab:consulting-agencies}
\begin{tabularx}{\linewidth}{ p{4em} X X X}
\toprule
\textbf{Consulting Agencies} & \textbf{ORCAA} & \textbf{Eticas} & \textbf{Babl AI} \\
\midrule
\hline


\textbf{Motivation} & regulatory compliance, performance testing, measure and mitigate disparate performance & help companies comply with legal requirements, security and data protection, fairness, bias and model accuracy & bias assessment, risk and impact assessment \\
\hline

\textbf{Target} & predictive scoring systems, hiring algorithms, healthcare AI, AI platforms, ADS, FRT & risk assessment algorithms; social media platforms such as tiktok and youtube; ride hailing apps, such as uber, cabify and bolt; algorithms used by governments and companies & their clients include universities, small AI companies, startups like Proctorio\\
\hline

\textbf{Types of harm} & discrimination, bias, measure gender and race/ethnicity bias, assess for regulatory compliance & algorithmic systems (their functioning), to detect anomalies or practices that could be unfair towards protected groups or society & bias, fairness, effectiveness, transparency, best policies and standards, privacy, compliance \\
\hline

\textbf{Institutional context} & The agency assists tech giants and governments, while its audits prioritize the concerns of marginalized communities & The consulting agency champions fairness for protected groups, yet also safeguards corporations from financial and reputational risks. & Ultimately, they are consultants that work with/for clients to help orgs build trust with stakeholders. \\
\hline

\textbf{Methods} & internal tool (Ethical Matrix framework), review documents & data management planning, encryption, ethics due diligence and vetting & in-house developed set of criteria used to conduct conduct bias testing, review documentations, interact with stakeholders\\
\hline

\textbf{Impact} & some influence (albeit indirect) on policy (on the White House AI ethics blueprint) & not clear & none stated \\

\bottomrule
\end{tabularx}
\end{table*}
\begin{table*}
\centering
\caption{Corporate Audits}
\label{tab:corporate-audits}
\begin{tabularx}{\linewidth}{ p{4em} X X X}
\toprule
\textbf{Corporate Audits} & \textbf{Google} & \textbf{Facebook} \\
\midrule


\textbf{Motivation} & identify potential issues relating to celebrity recognition products & help the company identify, prioritise, and implement improvements in accordance with civil rights "at the behest of the civil rights community and members of the US congress" \\
\hline

\textbf{Target} & Google’s celebrity recognition API
 & Facebook \\
\hline

\textbf{Types of harm} & human rights *impacts* & racial injustice, voter suppression, hate speech, algorithmic bias \\
\hline

\textbf{Methods} & consultation with stakeholders, dialogue with expert resources & interview with hundreds of civil rights orgs and advocates and members of congress \\
\hline

\textbf{Impact} & incorporation of some of the findings into product design & facebook committed to implement (no information if these changes were in fact implemented) a new advertising system in accordance with civil rights \\

\bottomrule
\end{tabularx}
\end{table*}
\begin{table*}
\centering
\caption{Journalism}
\label{tab:journalism}
\begin{tabularx}{\linewidth}{ p{7em} X X}
\toprule
\textbf{Journalism} & \textbf{The MarkUp} & \textbf{Propublica} \\
\midrule


\textbf{Motivation} & Accountability, expose disparities injustice and illegal practice & accountability, expose disparities in performance, privacy violations, discrimination in deployed systems \\
\hline

\textbf{Target} & arge corporations/institutions, social media platforms, algorithms, content moderators
 & Gov't ADS, VLOPs (Facebook advertising, Amazon product placement) public programs with digital sites \\
\hline

\textbf{Types of harm} & Disparities, fraud, discrimination, legal compliance, privacy breach & discrimination, privacy harms, cultural genocide, injustice \\
\hline

\textbf{Institutional context} & Targeting mostly tech firms/platforms, on behalf of general public; funded by foundations and private donors & Targeting public agencies and private companies, on behalf of general public; funded by foundations and private donors \\
\hline

\textbf{Methods} & Inhouse audit tools, interview, data donation, reviewing documents& scraping, data donation, API exploitation, evaluation, interviews, qualitative evidence \\
\hline

\textbf{Impacts} & high impact resulting in legislative and systemic (in the US)
 & High impact, legislative action, company policy changes, COMPAS a canonical work that set president for academic research \\

\bottomrule
\end{tabularx}
\end{table*}
\begin{table*}
\centering
\caption{Law Firms}
\label{tab:law-firms}
\begin{tabularx}{\linewidth}{ p{4em} X X X}
\toprule
\textbf{Law firms} & \textbf{BNH.AI} & \textbf{Foxglove} & \textbf{AWO} \\
\midrule

\textbf{Audit type} & information not available & Case study. algorithms, social media platforms, living conditions of tech workers & evaluative reviews of general industries or technologies, case studies \\
\hline

\textbf{Audit Goal} & compliance with standards & keeping tech giants and governments accountable & evaluate/design governance, protect data rights \\
\hline

\textbf{Audit target} & models and data & social media platforms, governments, content moderator, warehouse workers, gig-workers & VLOPs, private companies, FRT \\
\hline

\textbf{Type of harm} & legal compliance/liability/legal defence & ustice, disparate performance, enforcing legal requirements & privacy/data rights/surveillance violations, safety, digital manipulation/exploitation \\
\hline

\textbf{Methods} & information not available & legal compliance, anecdotal evidence & documentation, critical commentary, policy development, legal/compliance research \\
\hline

\textbf{Impact} & information not available & Sforced disclosure of secret contract between tech corps and governments, stopped the UK Home Office use of visa-streaming algo, reversed Ofqual's A level grading algorithm & litigation, change to govt's and corporate policy/strategy \\
\hline

\textbf{Power Analysis} & corporate/for profit, confidential and privileged & non-profit & corporate/for profit, consulting for both corporate and public agencies \\

\bottomrule
\end{tabularx}
\end{table*}
\begin{table*}
\centering
\caption{Regulators}
\label{tab:regulators}
\begin{tabularx}{\linewidth}{ p{4em} X X X}
\toprule
\textbf{Regulators} & \textbf{ICO} & \textbf{NIST} \\
\midrule
\textbf{General Objectives} & Uphold information rights. Enforcing legal requirements. Regulatory interventions through guidance, enforcement notice, and issuing monetary penalties. & Quality assurance, develop best standards and regulatory practices \\
\hline

\textbf{Audit type} & Case studies & Meta-commentary \\
\hline

\textbf{Audit Goal} & Investigate and enforce regulations & Establish standards \\
\hline

\textbf{Audit target} & Data, data sharing and management documents
 & FRT, ADS \\
\hline

\textbf{Type of harm} & Legal compliance, privacy breach, data management & Racial disparities, physical safety, functionality, privacy \\
\hline

\textbf{Methods} & Interview, testing, reviewing documents & Data trust, accuracy evaluation, benchmarking \\
\hline

\textbf{Impact} & Significant monetary penalties issued & Significant contribution on standardisation \\
\hline

\textbf{Power Analysis} & Targets companies and gov’t agencies; has authority to enforce laws with fines & Targets companies and gov’t agencies; establishes standards but no enforcement authority \\

\bottomrule
\end{tabularx}
\end{table*}
\begin{table*}
\centering
\caption{Civil Society}
\label{tab:civil-society}
\begin{tabularx}{\linewidth}{ p{7em} X X X X X X}
\toprule
\textbf{Civil Society} & \textbf{EFF} & \textbf{RLL} & \textbf{The Citizen Lab} & \textbf{MTM} & \textbf{Ada Lovelace} & \textbf{ACLU}\\
\midrule
\textbf{Motivation} & accountability, non-profit and independent & accountability, non-profit and independent & accountability & accountability & accountability & accountability \\
\hline


\hline

\textbf{Target} & ADS, online platforms, large corporations & ADS, lie detectors, border patrolling drones & ADS, apps, code, websites, FRT, surveillance tech, hardware (phones and other devices used by politicians and activists), software & biometric data, digital ID systems, border surveillance vendors, surveillance drones, FRT, iris scan data, lie detector tech, thermal cameras, algorithmic motion detectors, AI powered satellites, ankle monitors and GPS tags, voice recognition tech & AI and data-driven systems generally, biometric data (facial recognition) and healthcare in particular & ADS, platform advertising, facial recognition tech, medical algorithms, online advertising (particularly Facebook), welfare programs, insurance pricing, redlining algorithms \\
\hline

\textbf{Types of harm} & security vulnerabilities, fair use, user rights, privacy, free speech, resisting surveillance, encryption, consumer protection & disparities in refugee claim recognition rates, human rights violations, digital rights violations, xenophobic and racial discrimination, privacy & privacy, security, transparency, surveillance, spyware, censorship, freedom of expression & racial discrimination, border surveillance, xenophobia, human rights violations & privacy harms, due process harms, disparate impact & fallacy of functionality, cultural hegemony, hate speech / toxicity \\
\hline

\textbf{Methods} & litigation, policy analysis, grassroot activism, technology development & interviews with refugees, data collection, analysis and interactive visualisation from the Immigration and Refugee Board (IRB) through Access to Information Request, documentations of deportation and refused refugee claims, film making  & in-house tools, forensic analysis of devices, interviews  & document and archive migration tech, interviews with people crossing borders, investigative analysis, photography & participatory methods, including opinion polling; also policy research and policy development & quantitative evaluation of operational systems (but not necessarily as rigorous as academic papers; may be simple metrics); data acquired through privileged access, public records requests, scraping \\
\hline

\textbf{Impact} & high impact. Sued and won a case for First Amendment protection for software code for privacy protection, class-action lawsuit that resulted in removing Sony BMG off the market, won a rulings against the US government’s attempt to track location of mobile phone users, won Proctorio lawsuit on behalf of a student & some impact around advocacy
 & significant media coverage and attention to some investigations & some impact. TUI stopped deportation flights as a result of MTM’s campaigns &   &   \\

\bottomrule
\end{tabularx}
\end{table*}
\clearpage

\end{document}